\documentclass[]{aa}
\usepackage[varg]{txfonts}
\usepackage{graphicx}
\begin{document}

\title{1000 au Exterior Arcs Connected to the Protoplanetary Disk around \object{HL~Tau}}
\author{Hsi-Wei Yen\inst{\ref{inst1}}, Shigehisa Takakuwa\inst{\ref{inst2},\ref{inst3}}, You-Hua Chu\inst{\ref{inst3}}, Naomi Hirano\inst{\ref{inst3}}, Paul T. P. Ho\inst{\ref{inst3},\ref{inst4}}, Kazuhiro D. Kanagawa\inst{\ref{inst5}}, Chin-Fei Lee\inst{\ref{inst3}}, Hauyu Baobab Liu\inst{\ref{inst1}}, Sheng-Yuan Liu\inst{\ref{inst3}}, Tomoaki Matsumoto\inst{\ref{inst6}}, Satoki Matsushita\inst{\ref{inst3}}, Takayuki Muto\inst{\ref{inst7}}, Kazuya Saigo\inst{\ref{inst8}}, Ya-Wen Tang\inst{\ref{inst3}}, Alfonso Trejo\inst{\ref{inst3}}, Chun-Ju Wu\inst{\ref{inst9},\ref{inst3}}}

\institute{European Southern Observatory (ESO), Karl-Schwarzschild-Str. 2, D-85748 Garching, Germany \email{hyen@eso.org}\label{inst1} 
\and
Department of Physics and Astronomy, Graduate School of Science and Engineering, Kagoshima University, 1-21-35 Korimoto, Kagoshima, Kagoshima 890-0065, Japan\label{inst2}
\and 
Academia Sinica Institute of Astronomy and Astrophysics, P.O. Box 23-141, Taipei 10617, Taiwan\label{inst3}
\and
East Asian Observatory, 660 N. A'ohoku Place, University Park, Hilo, HI 96720, USA\label{inst4}
\and
Institute of Physics and CASA*, Faculty of Mathematics and Physics, University of Szczecin, Wielkopolska 15, 70-451 Szczecin, Poland\label{inst5}
\and
Faculty of Sustainability Studies, Hosei University, Chiyoda-ku, Tokyo 102-8160, Japan\label{inst6}
\and
Division of Liberal Arts, Kogakuin University, 1-24-2 Nishi-Shinjuku, Shinjuku-ku, Tokyo 163-8677, Japan\label{inst7}
\and
ALMA Project Office, National Astronomical Observatory of Japan, Osawa 2-21-1, Mitaka, Tokyo 181-8588, Japan\label{inst8}
\and
Department of Physics, National Taiwan University, Taipei 10617, Taiwan\label{inst9}
}
\date{Received date / Accepted date}

\abstract
{}
{The protoplanetary disk around \object{HL~Tau} is so far the youngest candidate of planet formation, and it is still embedded in a protostellar envelope with a size of thousands of au. In this work, we study the gas kinematics in the envelope and its possible influence on the embedded disk.} 
{We present our new ALMA cycle 3 observational results of \object{HL~Tau} in the $^{13}$CO (2--1) and C$^{18}$O (2--1) emission at resolutions of 0\farcs8 (110 au), and we compare the observed velocity pattern with models of different kinds of gas motions.}
{Both the $^{13}$CO and C$^{18}$O emission lines show a central compact component with a size of 2$\arcsec$ (280 au), which traces the protoplanetary disk. 
The disk is clearly resolved and shows a Keplerian motion, from which the protostellar mass of \object{HL~Tau} is estimated to be 1.8$\pm$0.3 $M_\sun$, assuming the inclination angle of the disk to be 47$\degr$ from the plane of the sky. 
The $^{13}$CO emission shows two arc structures with sizes of 1000--2000 au and masses of 3 $\times$ 10$^{-3}$ $M_\sun$ connected to the central disk. 
One is blueshifted and stretches from the northeast to the northwest, and the other is redshifted and stretches from the southwest to the southeast.
We find that simple kinematical models of infalling and (counter-)rotating flattened envelopes cannot fully explain the observed velocity patterns in the arc structures. 
The gas kinematics of the arc structures can be better explained with three-dimensional infalling or outflowing motions. 
Nevertheless, the observed velocity in the northwestern part of the blueshifted arc structure is $\sim$60--70\% higher than the expected free-fall velocity. 
We discuss two possible origins of the arc structures: (1) infalling flows externally compressed by an expanding shell driven by \object{XZ~Tau} and (2) outflowing gas clumps caused by gravitational instabilities in the protoplanetary disk around \object{HL~Tau}.}
{}


\titlerunning{arc Structures Connected to the HL Tau Disk}
\authorrunning{H.-W. Yen et al.}

\maketitle

\section{Introduction}
Observations in (sub-)millimeter continuum at spatial resolutions of a few au and high-contrast imaging at infrared have revealed rings and gaps with widths of few au to tens of au in several protoplanetary disks \citep{Akiyama15, Momose15, Rapson15, Isella16, Andrews16, Ginski16, Konishi16, Perrot16, Plas17}. 
The presence of these gaps suggests decreases in surface density or changes in dust properties at the location of these gaps \citep{Takahashi14, Dipierro15, Dong15, Dong16, Kanagawa15, Kanagawa16, Tamayo15, Zhang15, Jin16, Okuzumi16, Dong17}. 
Hints of decreases in gas density coincident with the dust continuum gaps have also been observed in protoplanetary disks \citep{Yen16, Isella16}, 
suggestive of actual deficits in material at these gaps. 
Theoretical studies show that embedded gas giant planets can carve such gaps in protoplanetary disks \citep[e.g.,][]{Dong15, Dong16, Kanagawa15, Kanagawa16, Dong17}. 
Thus, these protoplanetary disks with rings and gaps are considered as candidates of on-going planet formation. 

\object{HL~Tau} (J2000 04$^{\rm h}$31$^{\rm m}$38$.\!\!^{\rm s}$43 +18$^{\rm d}$13$^{\rm m}$57$.\!\!^{\rm s}$2; \citet{ALMA15}) is a protostar with a Class I--II spectral energy distribution \citep{Men'shchikov99} and is located in the Taurus star-forming region at a distance of 140 pc \citep{Kenyon94, Loinard13}. 
A series of rings and gaps were observed in its protoplanetary disk in the (sub-)millimeter continuum emission with Atacama Large Millimeter/Submillimeter Array (ALMA), making \object{HL~Tau} so far the youngest candidate of planet formation \citep{ALMA15, Akiyama16}. 
Direct imaging in the $L^\prime$ band with the Large Binocular Telescope Interferometer did not find any point sources in the protoplanetary disk around \object{HL~Tau} and put an upper limit on the mass of possible planets to be 10--15 $M_{\rm jup}$ \citep{Testi15}.
The protoplanetary disk around \object{HL~Tau} has an inclination angle of 47$\degr$ from the plane of the sky and a position angle (PA) of 138$\degr$ \citep{ALMA15} and is embedded in an elongated molecular cloud with a length of $\sim$0.05 pc \citep{Welch00}. 
Observations in the $^{13}$CO (1--0) emission at an angular resolution of 5$\arcsec$ with the Nobeyama Millimeter Array show a flattened envelope with signs of infalling and rotational motions on a scale of 2000 au around \object{HL~Tau} \citep{Hayashi93}. 
Later observations in the $^{13}$CO (1--0) emission with PdBI and combination of BIMA and the NRAO 12m telescope at angular resolution of 3$\arcsec$--8$\arcsec$ reveal complex velocity structures on a scale of hundreds to thousands of au \citep{Cabrit96, Welch00}. 
Thus, the gas motions in the envelope around \object{HL~Tau} are still not clear. 
\object{HL~Tau} is also associated with a bipolar molecular outflow \citep{Monin96, Lumbreras14, ALMA15, Klaassen16} and optical and infrared jets \citep{Takami07, Hayashi09, Anglada07, Movsessian12}. 
The jets associated with \object{HL~Tau} are moving at tangential velocities of 60--160 km s$^{-1}$ with an inclination angle of 40$\degr$--60$\degr$ and a PA of 35$\degr$--47$\degr$, measured from the proper motion of the jets \citep{Anglada07, Movsessian12}.
The jet axis is perpendicular to the protoplanetary disk.
Therefore, 
different from other candidates of planet formation, 
\object{HL~Tau} is still surrounded by an protostellar envelope and is likely in the active mass accretion phase. 

Dynamical infall from protostellar envelopes onto protoplanetary disks can produce accretion shocks in the disks \citep[e.g.,][]{Yorke99}. 
This can change the chemical status of protoplanetary disks and possibly affect the subsequent chemical evolution of disks \citep[e.g.,][]{Visser09, Visser11}. 
Signs of accretion shocks have been observed in disks around Class 0 and I protostars, such as \object{L1489~IRS} \citep{Yen14}, \object{L1527} \citep{Ohashi14, Sakai14a, Sakai14b}, and \object{TMC-1A} \citep{Sakai16}. 
Changes in chemical compositions have also been observed in the transitional region between the infalling envelope and the central disk in \object{L1527} \citep{Sakai14a, Sakai14b} and \object{IRAS~16293$-$2422} \citep{Oya16}.
In addition, 
as a protoplanetary disk continues to accrete mass from its surrounding envelope, 
it can become massive and gravitational unstable \citep[e.g.,][]{Machida10, Vorobyov10}. 
The gravitational instability can cause fragmentation and form gas clumps in disks, which can be the progenitors of planets \citep[e.g.,][]{Vorobyov11, Zhu12}, 
and it can also trigger mass ejection from disks \citep[e.g.,][]{Vorobyov16}. 
These influences of protostellar envelopes on embedded protoplanetary disks are not well understood observationally. 

The presence of those gaps and rings in the protoplanetary disk around \object{HL~Tau} suggests planet formation may occur when protoplanetary disks are still embedded in protostellar envelopes. 
To observationally investigate possible influences of protostellar envelopes on planet-forming disks, 
we have conducted ALMA observations in $^{13}$CO (2--1; 220.398684 GHz) and C$^{18}$O (2--1; 219.560358 GHz) at an angular resolution of 0\farcs8 (110 au) toward the candidate disk of planet formation around \object{HL~Tau}. 
Our ALMA observations allow us to resolve the Keplerian rotation of the embedded protoplanetary disk and distinguish the gas motions in the surrounding envelope, 
and thus to study their relation.
This paper is organized as follows: Section 2 describes the details of the observations. Section 3 presents the observational results in the $^{13}$CO and C$^{18}$O emission. Section 4 presents our analysis on the disk rotation traced by the $^{13}$CO and C$^{18}$O emission and on the kinematics of the surrounding structures connected to the disk observed in the $^{13}$CO emission. Section 5 discusses the origins of the observed structures and kinematics in the $^{13}$CO emission. 

\section{Observations}
ALMA cycle-3 observations toward \object{HL~Tau} were conducted on 22 March, 2016 with 40 antennas. 
\object{HL~Tau} was observed with 7 pointing mosaic, and the total integration time on \object{HL~Tau} is 30 min.
J0510+1800 with a flux of 2.38 Jy at 225.7 GHz was observed as bandpass and flux calibrators, and J0431+2037 with a flux of 65 mJy at 225.6 GHz as a phase calibrators. 
The baseline length ranges from 14 m to 384 m, 
so the maximum recoverable scale of our observations is 12$\arcsec$ (1700 au). 
The typical absolute flux uncertainty of ALMA observations at 1 mm wavelength is 10\%. 
The spectral setup of our ALMA observations contained five spectral windows.
Two spectral windows, each with a bandwidth of 2 GHz, were assigned to the 1.3 mm continuum. 
One spectral window with a bandwidth of 468.8 MHz and a channel width of 122 kHz was set to the $^{13}$CO (2--1) line, one with a bandwidth of 234.4 MHz and a channel width of 244 kHz to the C$^{18}$O (2--1) line, and one with a bandwidth of 234.4 MHz and a channel width of 488 kHz to the SO (5$_6$--4$_5$) line.
The raw visibility data were calibrated using the standard reduction script for the cycle-3 data, which uses tasks in Common Astronomy Software Applications (CASA; \citet{McMullin07}) of version 4.6, and without self-calibration. 
We additionally flagged the data in the spectral window of the $^{13}$CO line in the field 4 at time from 21:40:14 to 21:40:16 manually, because artificial stripes appear in images produced with uniform weighting. 
The image cubes of the $^{13}$CO (2--1) and C$^{18}$O (2--1) lines were generated with the uniform weighting and cleaned with the CASA task ``clean'' at a velocity resolution of 0.34 km s$^{-1}$. 
The angular resolution and the noise level are 0\farcs77 $\times$ 0\farcs75 and 7 mJy Beam$^{-1}$ per channel in the $^{13}$CO image, and 0\farcs77 $\times$ 0\farcs74 and 6.7 mJy Beam$^{-1}$ per channel in the C$^{18}$O image, respectively.

\section{Results}

\begin{figure*}
\centering
\includegraphics[width=17cm]{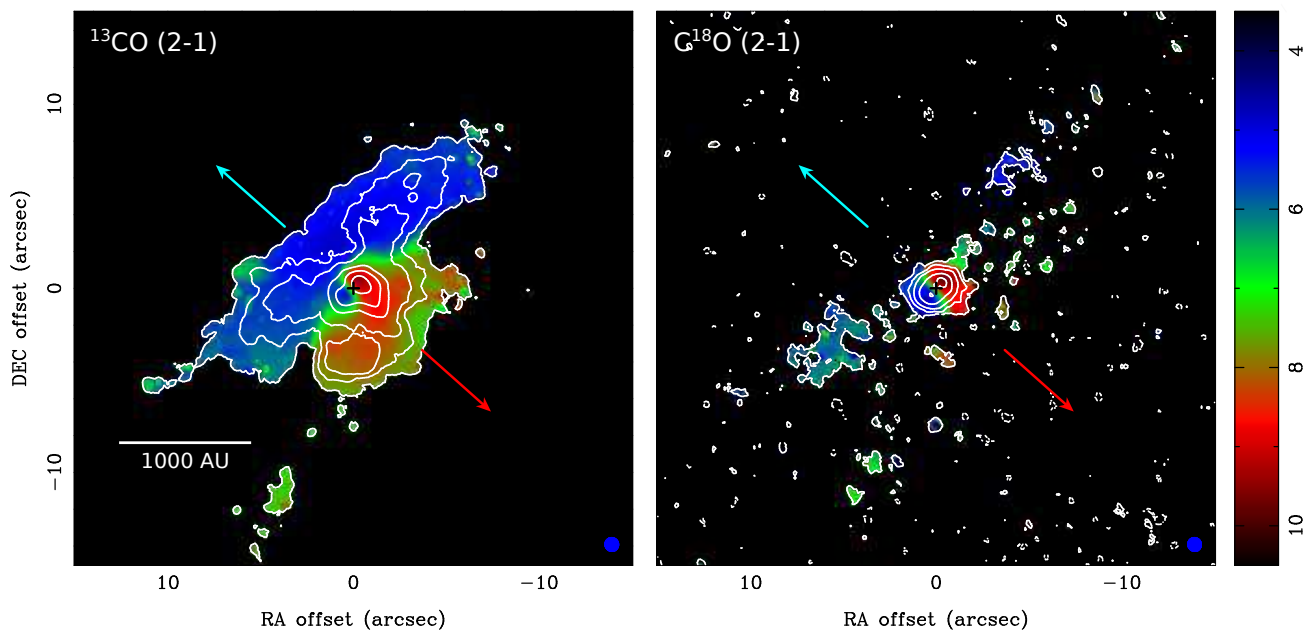}
\caption{Moment 0 maps (contours) overlaid on moment 1  maps (color; in units of km s$^{-1}$ in the LSR frame) of the C$^{13}$O (2--1; left) and C$^{18}$O (2--1; right) emission in \object{HL~Tau} obtained with our ALMA observations. Crosses denote the position of \object{HL~Tau}. Blue and red arrows show the directions of the blue- and redshifted outflows. Blue filled ellipses present the sizes of the synthesized beams. In the $^{13}$CO map, contour levels are 5\%, 10\%, 15\%, 30\%, 50\%, and 80\% of the peak intensity, where the 5\% level corresponds to 4.5$\sigma$ (1$\sigma$ = 13.7 mJy Beam$^{-1}$ km s$^{-1}$). Those in the C$^{18}$O map are 5\%, 15\%, 30\%, 60\%, and 90\% of the peak intensity, where the 5\% level corresponds to 2$\sigma$ (1$\sigma$ = 11.6 mJy Beam$^{-1}$ km s$^{-1}$).}\label{mom}
\end{figure*}

Figure \ref{mom} presents the total integrated intensity (moment 0) maps overlaid on the intensity-weighted mean velocity (moment 1) maps of the $^{13}$CO (2--1) and C$^{18}$O (2--1) emission. 
The C$^{18}$O emission primarily traces a compact component with an apparent size of $\sim$2$\arcsec$ (280 au) around \object{HL~Tau}, 
and it is oriented along the northwest--southeast direction with a PA of $\sim$125$\degr$. 
This orientation is consistent with that of the protoplanetary disk around \object{HL~Tau} observed with ALMA  at an angular resolution of 0\farcs03 within 15$\degr$ \citep{ALMA15, Akiyama15}.
The C$^{18}$O central component exhibits a velocity gradient along its major axis, where the northwestern part is redshifted and the southeastern part is blueshifted. 
The direction of this velocity gradient is the same as the Keplerian rotation of the protoplanetary disk around \object{HL~Tau} observed with ALMA in the HCO$^+$ (1--0) and CO (1--0) emission at angular resolutions of 0\farcs2--0\farcs3 \citep{Pinte16}. 
The $^{13}$CO emission also shows a similar central compact component, which is orientated and exhibits a velocity gradient along the northwest--southeast direction. 
Besides, in the $^{13}$CO emission, there are extended structures with sizes of $\sim$1000--3000 au linking the central component. 
The extended structures stretch toward the northwest and the southeast, and their elongations are perpendicular to the directions of the molecular outflows and the optical and infrared jets \citep{Monin96, Anglada07, Takami07, Movsessian07, Movsessian12, Hayashi09, Lumbreras14, ALMA15, Klaassen16}. 
A part of the extended structures is also detected in the C$^{18}$O emission, which appears as clumpy components located to the northwest, the southeast, and the south of the central component.
The extended structures exhibit an overall velocity gradient in the same direction as the molecular outflows associated with \object{HL~Tau} \citep{Monin96, Lumbreras14, ALMA15, Klaassen16}. 
This velocity gradient on a 1000 au scale is also observed in the $^{13}$CO (1--0) emission at lower resolutions of 3$\arcsec$--8$\arcsec$ \citep{Sargent87, Sargent91, Hayashi93, Cabrit96, Welch00}.

\begin{figure*}
\centering
\includegraphics[width=17cm]{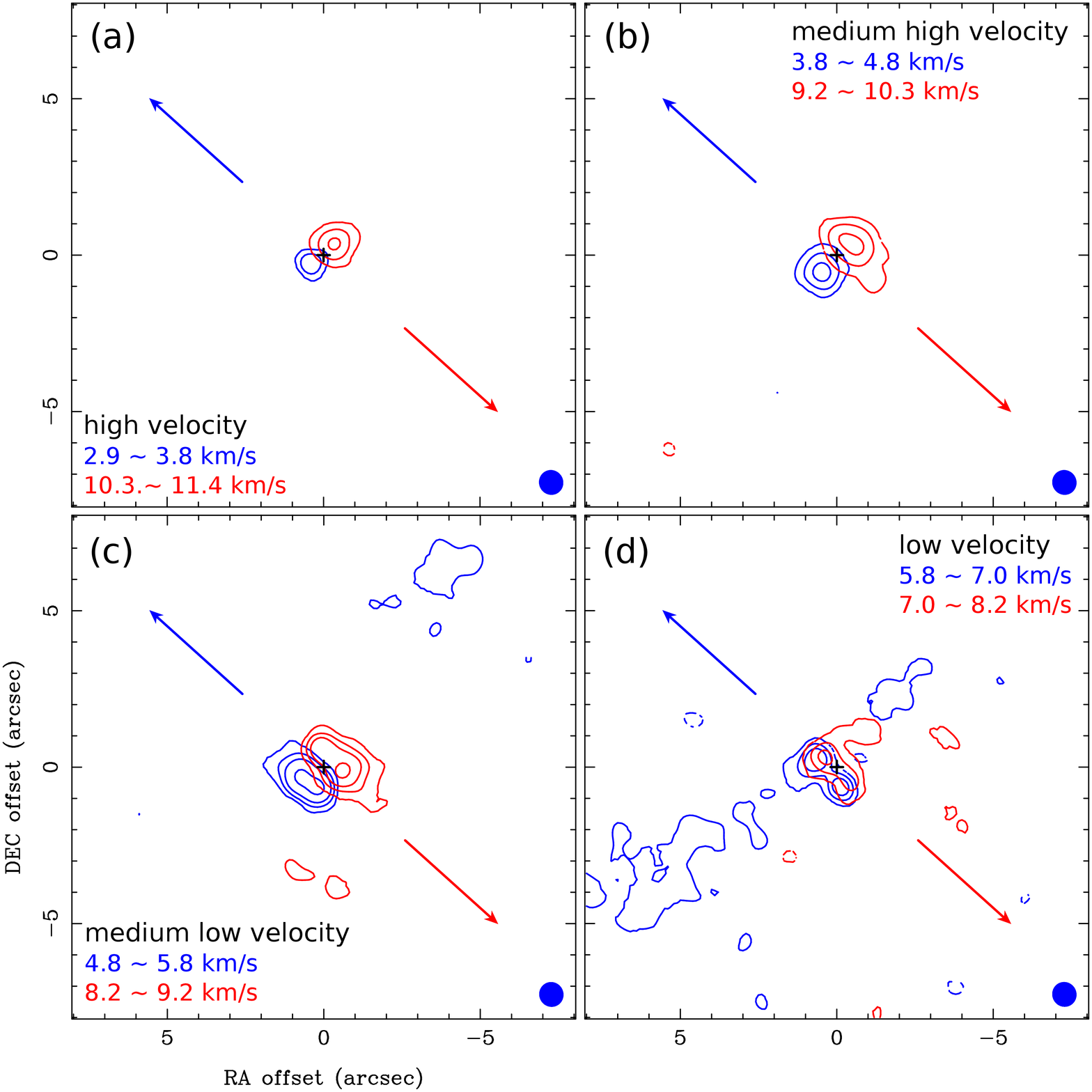}
\caption{Moment 0 maps of the C$^{18}$O emission integrated over four different velocity ranges, high velocities ($V_{\rm LSR}$ = 2.9--3.8 and 10.3--11.4 km s$^{-1}$), medium-high velocities ($V_{\rm LSR}$ = 3.8--4.8 and 9.2--10.3 km s$^{-1}$), medium-low velocities ($V_{\rm LSR}$ = 4.8--5.8 and 8.2--9.2 km s$^{-1}$), and low velocities ($V_{\rm LSR}$ = 5.8--7.1 and 7.1--8.2 km s$^{-1}$). Blue and red contours present the blue- and redshifted emission, respectively. Blue and red arrows show the directions of the blue- and redshifted outflows. Crosses denote the position of \object{HL~Tau}. Filled blue ellipses show the size of the synthesized beam. Contour levels are 5$\sigma$,10$\sigma$, and 15$\sigma$ in the high-velocity blueshifted map, are 5$\sigma$, 15$\sigma$, 30$\sigma$, 50$\sigma$, and 70$\sigma$ in the high-velocity redshifted map, are 5$\sigma$,15$\sigma$, 30$\sigma$, and 45$\sigma$ in the medium-high-velocity maps, 5$\sigma$,10$\sigma$,15$\sigma$, 25$\sigma$, 45$\sigma$, and 75$\sigma$ in the medium-low-velocity maps, and are 5$\sigma$, 10$\sigma$, 15$\sigma$, 25$\sigma$, 45$\sigma$, and 75$\sigma$ in the low-velocity maps. The noise level in the high-velocity redshifted map is 5.1 mJy Beam$^{-1}$ km s$^{-1}$, and those in the other maps are all 3.9 mJy Beam$^{-1}$ km s$^{-1}$.}\label{momc18o}
\end{figure*}

\begin{figure*}
\centering
\includegraphics[width=17cm]{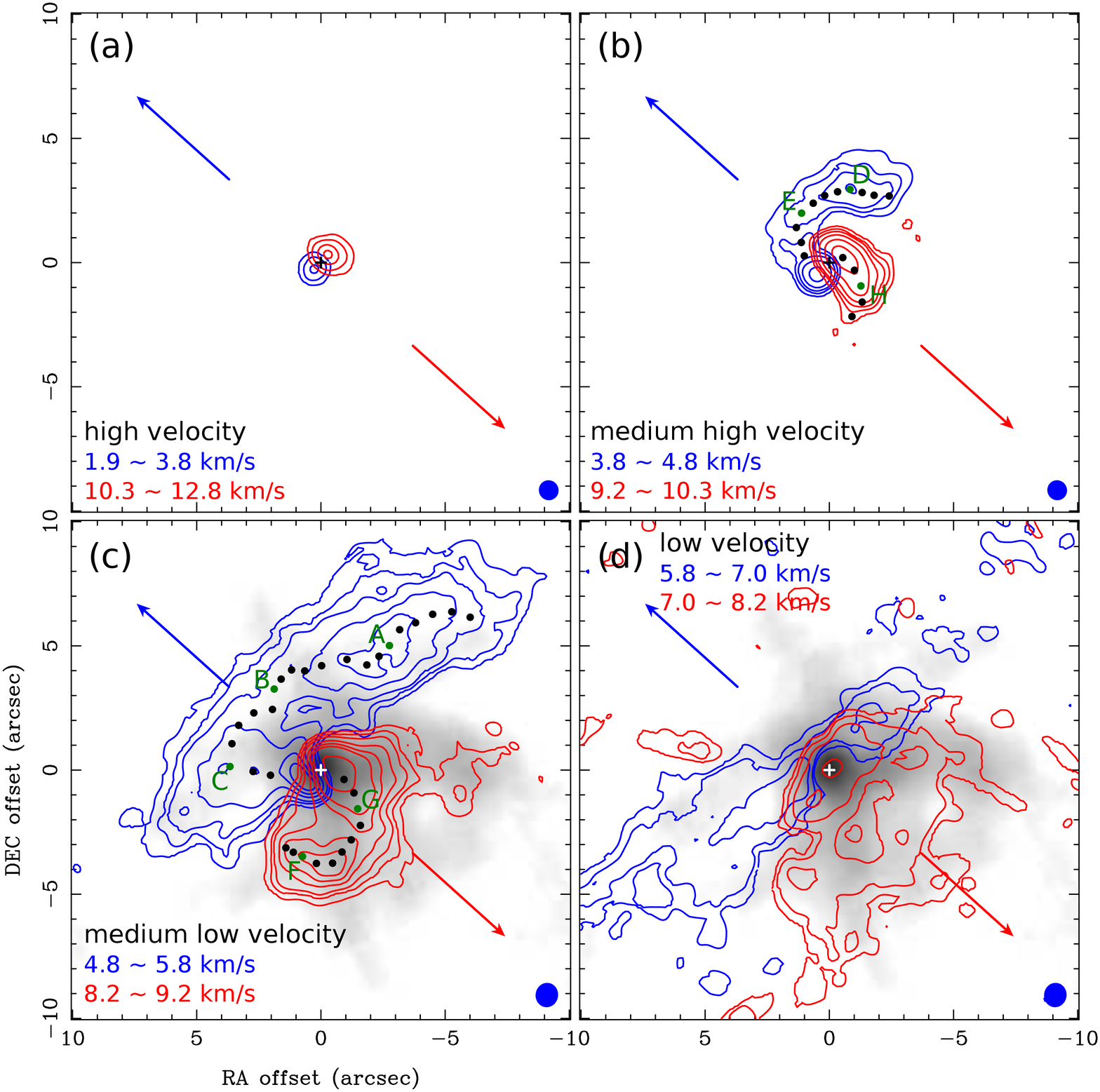}
\caption{Same as Fig.~\ref{momc18o} but for the $^{13}$CO emission, and the high-velocity range for the $^{13}$CO emission is $V_{\rm LSR}$ = 1.9--3.8 and 10.3--12.8 km s$^{-1}$. Black and green dots delineate the ridges of the contour lines, showing the arc structures. The ridges of the contour lines were measured from the peak positions of the intensity profiles along the transverse directions of the arc structures. Green dots with label A--H show the positions of the extracted spectra shown in Fig.~\ref{13cospec}. Grey scales in (c) and (d) present the moment 0 map of the CO (1--0) emission observed with ALMA, and the map is obtained from the data released by \citet{ALMA15}. Contour levels are 5$\sigma$,15$\sigma$, and 30$\sigma$ in the high-velocity blueshifted map, are 5$\sigma$, 15$\sigma$, 35$\sigma$, and 55$\sigma$ in the high-velocity redshifted map, are 5$\sigma$,15$\sigma$, 25$\sigma$, 45$\sigma$, and 75$\sigma$ in the medium-high-velocity maps, 5$\sigma$,10$\sigma$,15$\sigma$, 25$\sigma$, 35$\sigma$, 45$\sigma$, and 75$\sigma$ in the medium-low-velocity maps, and are 5$\sigma$, 10$\sigma$, 15$\sigma$, 20$\sigma$, and 60$\sigma$ in the low-velocity maps. The noise level in the high-velocity blue- and redshifted map is 5.8 and 7.1 mJy Beam$^{-1}$ km s$^{-1}$, respectively, and those in the other maps are all 4.1 mJy Beam$^{-1}$ km s$^{-1}$. There is significant missing flux in the velocity ranges of (c) and (d), as shown in Fig.~\ref{chan13co} and \ref{chan13co2}. The negative contours are not plotted in (c) and (d) for the clarity of the figures.}\label{mom13co}
\end{figure*}

Figure \ref{momc18o} and \ref{mom13co} presents the moment 0 maps of the C$^{18}$O and $^{13}$CO emission integrated over the four different velocity ranges, respectively. 
The integrated velocity ranges are high velocities ($V_{\rm LSR}$ = 2.9--3.8 and 10.3--11.4 km s$^{-1}$ for C$^{18}$O and $V_{\rm LSR}$ = 1.9--3.8 and 10.3--12.8 km s$^{-1}$ for $^{13}$CO), medium-high velocities ($V_{\rm LSR}$ = 3.8--4.8 and 9.2--10.3 km s$^{-1}$), medium-low velocities ($V_{\rm LSR}$ = 4.8--5.8 and 8.2--9.2 km s$^{-1}$), and low velocities ($V_{\rm LSR}$ = 5.8--7.1 and 7.1--8.2 km s$^{-1}$). 
The systemic velocity ($V_{\rm sys}$) is measured to be $V_{\rm LSR}$ = 7.04 km s$^{-1}$ (Sect. \ref{kep}).
The high-velocity blue- and redshifted C$^{18}$O components are compact with sizes of 0\farcs3$\pm$0\farcs04 (42$\pm$6 au) and 0\farcs44$\pm$0\farcs02 (62$\pm$3 au), and are located in the southeast and the northwest to \object{HL~Tau} with a distance of 0\farcs5$\pm$0\farcs04 (70$\pm$6 au), respectively. 
At the medium-high velocities, the sizes of the blue- and redshifted C$^{18}$O components increase to 0\farcs6--0\farcs9 (80--130 au), 
and the distance between their peak positions and \object{HL~Tau} increases to 0\farcs6--0\farcs7 (80--100 au). 
These results show that the C$^{18}$O emission at higher velocities is located closer to \object{HL~Tau}. 
Furthermore, the axis passing through the peak positions of the high-velocity blue- and redshifted components has a PA of 130\degr$\pm$3\degr, and that of medium-high-velocity components is 133\degr$\pm$1\degr. 
Therefore, these components are well aligned along the major axis of the protoplanetary disk around \object{HL~Tau}, whose PA is 138$\degr$ \citep{ALMA15}.
At the medium-low and low velocities, 
both the C$^{18}$O blue- and redshifted components are elongated along the disk's minor axis.
Besides, 
the low-velocity components are centered at the position of  \object{HL~Tau}.  
Hence, the velocity structures of the central C$^{18}$O emission is consistent with the observational signatures of Keplerian rotation \citep[e.g.,][]{Dutrey94}. 

\begin{figure*}
\centering
\includegraphics[width=17cm]{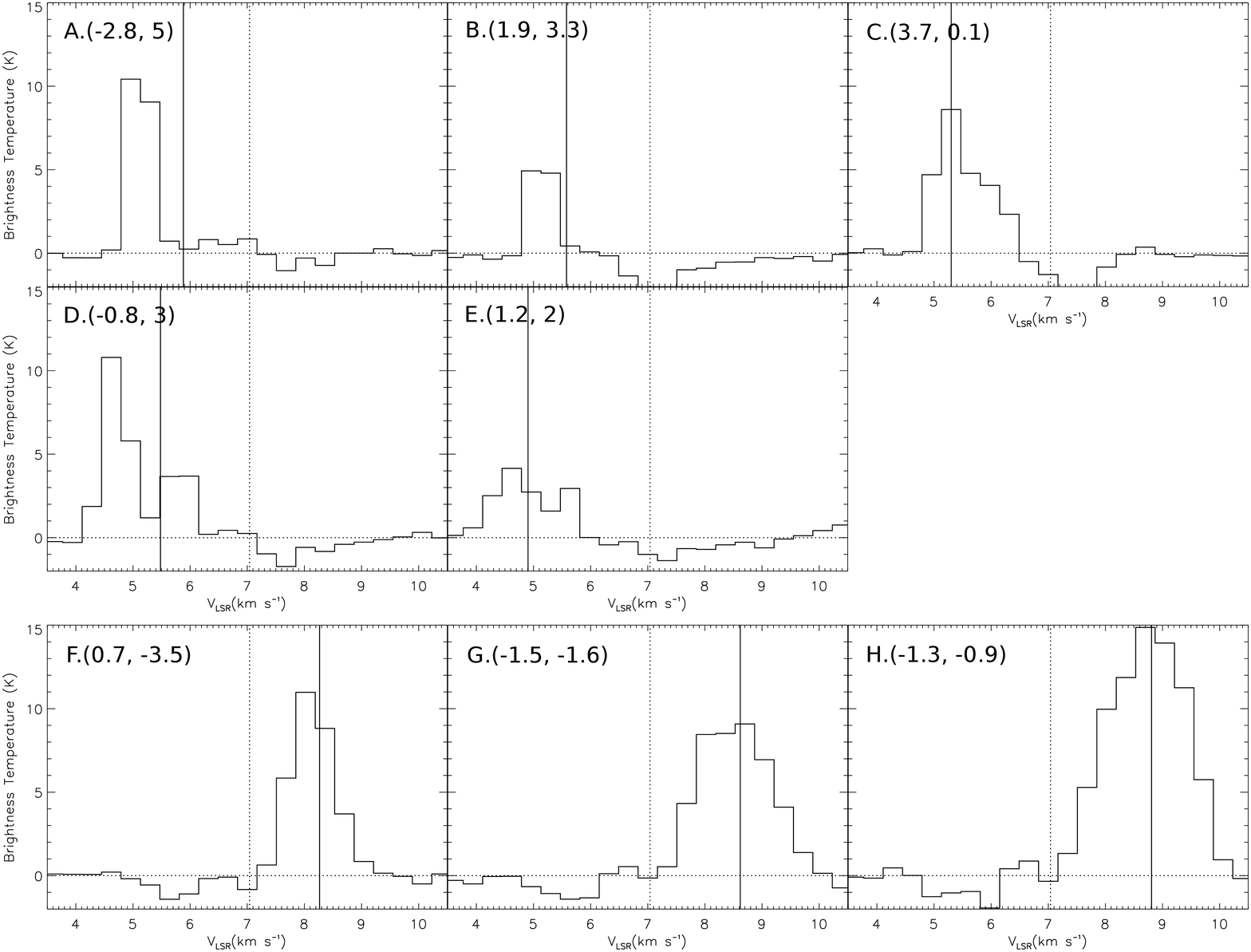}
\caption{Spectra of the $^{13}$CO emission along the arc structures. The positions A--H, where the spectra were extracted, relative to \object{HL~Tau} are labeled at the upper left corners in the panels in unit of arcsecond, and they are also labeled in Fig.~\ref{mom13co}. Vertical dotted lines denote the systemic velocity measured from the disk rotation. Vertical solid lines present the expected line-of-sight velocities computed from the model of the three-dimensional infalling and rotational motions (Sect. \ref{kin}).}\label{13cospec}
\end{figure*}

In the high-velocity $^{13}$CO  emission (Fig.~\ref{mom13co}a), the blue- and redshifted  components are compact with sizes of 0\farcs3--0\farcs5 (50--70 au) and located close to \object{HL~Tau} with a distance of 0\farcs4 (60 au), similar to the high-velocity C$^{18}$O emission. 
In addition, the axis passing through the peak positions of the high-velocity blue- and redshifted $^{13}$CO components is 135\degr$\pm$2\degr, so they are well aligned with the disk's major axis. 
Thus, the high-velocity $^{13}$CO  emission also exhibits the  observational signatures of dominant rotational motion.
The low-velocity blueshifted $^{13}$CO emission shows an extended and elongated structure along the northwest--southeast direction with a size of $\sim$20$\arcsec$ (2800 au; Fig.~\ref{mom13co}d). 
This extended structure is also partially detected in the C$^{18}$O emission (Fig.~\ref{momc18o}d).
Similar structures are also observed in the CO (1--0) emission with ALMA \citep{ALMA15}.
These structures could be related to the large-scale ambient gas. 
On the contrary, the low-velocity redshifted $^{13}$CO emission shows a bow-like structure with the apex located close to \object{HL~Tau}, and its morphology is similar to the wall of the outflow cavity observed in the CO emission with SMA and ALMA \citep{Lumbreras14, ALMA15, Klaassen16}.
Hence, the low-velocity redshifted $^{13}$CO emission likely traces the cavity wall of the redshifted outflow. 

At the medium velocities (Fig.~\ref{mom13co}b \& c), 
the $^{13}$CO emission exhibits the extended arc structures, as delineated by the ridges of their contour lines (dotted curves in Fig.~\ref{mom13co})$\footnotemark[1]$.  
The whole velocity channel maps of the $^{13}$CO emission in the medium-velocity ranges are shown in Appendix \ref{13coA}.
The blue- and redshifted arc structures connect to the central component from the northeast and the southwest, 
and they stretch toward the northwest and the southeast, respectively. 
Figure \ref{13cospec} presents the spectra of the $^{13}$CO emission at several positions along the arc structures. 
The line width of the blueshifted arc structures is 0.7 km s$^{-1}$, as seen at the offsets of A.($-2\farcs8$, 5\arcsec) and B.(1\farcs9, 3\farcs3). 
At the offsets of D.($-0\farcs8$, 3\arcsec) and E.(1\farcs2, 2\arcsec), there is a secondary component at $V_{\rm LSR} \sim 5.8$ km s$^{-1}$ in the spectra, and it is likely associated with the large-scale ambient gas as seen in Fig.~\ref{mom13co}d.  
The boarder line width at the offset of C.(3\farcs7, 0\farcs1) is most likely due to blending the blueshifted arc structure and the large-scale ambient gas. 
The spectra in the redshifted arc structure all show boarder line widths of 1.4--2.4 km s$^{-1}$ than those in the blueshifted one, as seen at the offsets of F.(0\farcs7, $-3\farcs5$), G.($-1\farcs5$, $-1\farcs6$), and H.($-1\farcs3$, $-0\farcs9$).
The redshifted arc structure is spatially coincident with the redshifted outflow observed at the low velocity (Fig.~\ref{mom13co}), 
and the two components are blended in the spectra. 
Thus, it is not straightforward to measure the line width of the redshifted arc structure. 
Nevertheless, the line width of the redshifted arc structure is most likely narrower than 1.4 km s$^{-1}$ based on the spectrum at the offset of F.(0\farcs7, $-3\farcs5$), 
where there is less contribution from the outflow. 

\footnotetext[1]{The ridges of the coutour lines are measured as follows. First we set a center, approximately the center of the curvature of the arc structures. Then we extracted a series of intensity profiles along the lines passing through that center in steps of PA of every 10$\degr$--20$\degr$. For each intensity profile, we fitted a Gaussian profile to measure the peak position. These peak positions denote the ridges of the countour lines shown in Fig.~\ref{mom13co}. The selected centers for the arc structures at the blueshifted medium-high velocity is ($-1\farcs5$, 0\farcs5), at the redshifted medium-high velocity is (1\farcs0, $-1\farcs5$), and at the redshifted medium-low velocity is (0\farcs5, $-1\farcs5$). The arc structure at the the blueshifted medium-low velocity is more extended, and two centers were selected, which are ($-4\farcs5$, $5\farcs0$) and ($1\farcs0$, $2\farcs0$), to extract the intensity profiles to cover the whole structure. We have done tests and confirmed that the derived ridges are not sensitive to the choices of the centers.}

The arc structures observed at the medium-high velocities are located closer to \object{HL~Tau} than those at the medium-low velocities, showing that the inner part has a higher velocity than the outer part (Fig.~\ref{mom13co}b \& c). 
Thus, there is a radial velocity gradient in each arc structure. 
Besides, the blue- and redshifted arc structures extend to the northwest and to the southeast, respectively.
That forms an overall velocity gradient from the northwest to the southeast on a 1000 au scale. 
The overall velocity gradient along the northwest--southeast direction on a 1000 au scale has been observed in the $^{13}$CO (1--0) emission at lower resolutions of 3$\arcsec$--8$\arcsec$ \citep{Sargent91, Hayashi93, Cabrit96, Welch00}. 
That was interpreted as Keplerian rotation around a protostar with a mass of 0.5--1 $M_\sun$ by \citet{Sargent91} and as sub-Keplerian rotation in a disk-like structure by \citet{Hayashi93}. 
However, the ALMA observations show that the direction of this velocity gradient on a 1000 au scale is opposite to that of the disk rotation on a 100 au scale, where the northwestern part is redshifted and the southeastern part is blueshifted. 

\section{Analysis}
\subsection{Keplerian Disk}\label{kep}

\begin{figure*}
\centering
\includegraphics[width=17cm]{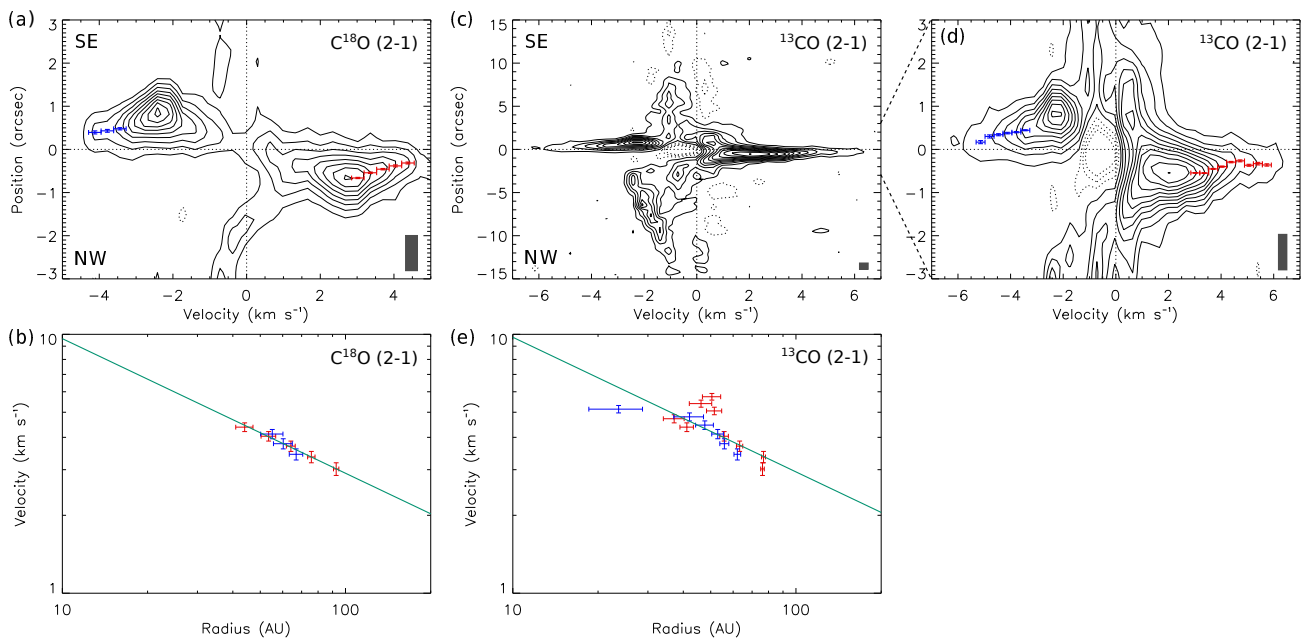}
\caption{Position--Velocity diagrams along the disk's major axis in the C$^{18}$O emission (a) and the $^{13}$CO emission (c). (d) is the zoom-in of (c). Blue and red data points show the measured peak positions in different velocity channels. Rotational profiles measured from the data points in (a) and (d) are shown in (b) and (e). Green solid lines show the best-fit power-law functions of the rotational profiles. Contour levels are from 3$\sigma$ in steps of 3$\sigma$ in panel (a), where 1$\sigma$ is 6.7 mJy Beam$^{-1}$, and those are from 3$\sigma$ in steps of 5$\sigma$ in panel (c) and (d), where 1$\sigma$ is 7 mJy Beam$^{-1}$. In the P--V diagrams, vertical and horizontal dotted lines denote the systemic velocity of $V_{\rm LSR} = 7.04$ km s$^{-1}$ and the stellar position, respectively, and grey rectangles at the bottom right corners show the angular and velocity resolutions.}\label{pvmaj}
\end{figure*}

\begin{figure*}
\centering
\includegraphics[width=17cm]{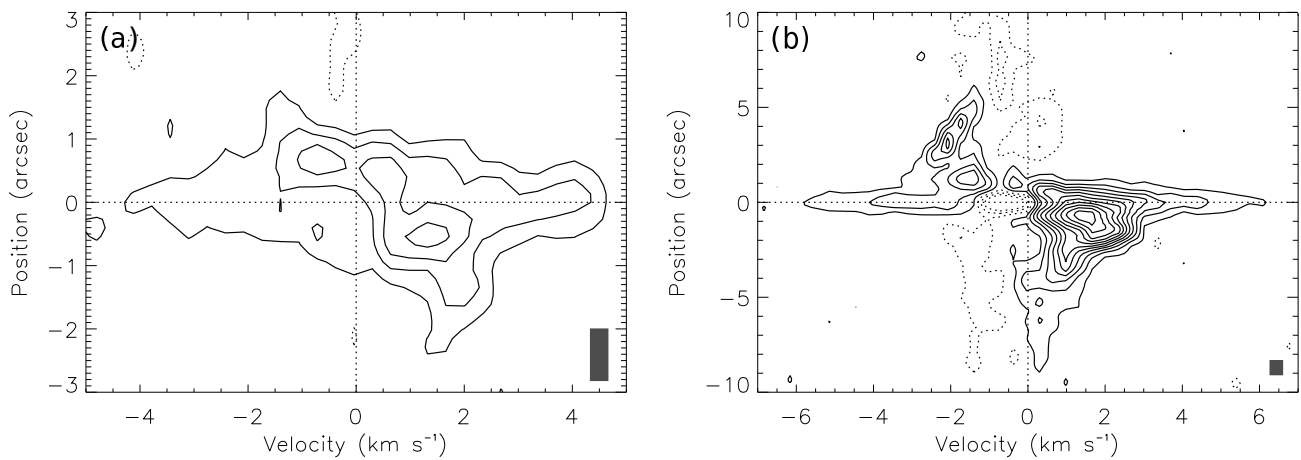}
\caption{Position--Velocity diagrams along the disk's minor axis in the C$^{18}$O emission (a) and the $^{13}$CO emission (b). Contour levels are from 3$\sigma$ in steps of 3$\sigma$ in panel (a) and are from 3$\sigma$ in steps of 5$\sigma$ in panel (b). Grey rectangles at the bottom right corners in the P--V diagrams show the angular and velocity resolutions.}\label{pvmin}
\end{figure*}

The C$^{18}$O and $^{13}$CO emission shows the observational signatures of rotation. 
To measure the rotational profiles in the C$^{18}$O and $^{13}$CO emission, 
we followed the method in \citet{Yen13}. 
We extracted position--velocity (P--V) diagrams along the disk's major axis (PA = 138\degr) and measured the peak positions in given velocity channels (Fig.~\ref{pvmaj}a, c, and d). 
The distances between the measured peak positions and \object{HL~Tau} are adopted as rotational radii ($R_{\rm rot}$), 
and the velocities at the velocity channels relative to the systemic velocity ($|V_{\rm LSR} - V_{\rm sys}|$) are adopted as rotational velocities ($V_{\rm rot}$).
We then fitted a power-law function to these data points, 
\begin{equation}
|V_{\rm LSR} - V_{\rm sys}| = V_{\rm rot}(R_0) \times (\frac{R_{\rm rot}}{R_0})^f {\rm\ km\ s^{-1}},
\end{equation}
where $R_0$ is the characteristic radius and is adopted to be 1$\arcsec$ (140 au) and $f$ is the power-law index of the rotational profile. 
Thus, there are three free parameters in the fitting, $V_{\rm rot}(R_0)$, $f$, and $V_{\rm sys}$.
This analysis was only applied to the C$^{18}$O and $^{13}$CO emission in the high-velocity ranges (Fig.~\ref{momc18o}a and \ref{mom13co}a), 
where there is no contamination from the extended structures.
Furthermore, 
there is no velocity gradient in the P--V diagram along the disk's minor axis at the high velocities (Fig.~\ref{pvmin}), 
suggesting that the gas motion is dominated by the rotation with least contamination from other motions in these velocity ranges. 
In constrast, at the medium and low velocities, there is a clear velocity gradient along the disk's minor axis in the P--V diagrams of the C$^{18}$O and $^{13}$CO emission, which could be due to an infalling motion or outflow. 

The rotational profile in the C$^{18}$O emission is measured to be $(2.44\pm0.07) \times (R_{\rm rot}/R_0)^{-0.52\pm0.04}$ km s$^{-1}$ with $V_{\rm sys} = 7.04\pm0.04$ km s$^{-1}$ (Fig.~\ref{pvmaj}b).
This rotational profile is consistent with Keplerian rotation ($f = -0.5$) within the uncertainty. 
The measured $V_{\rm sys}$ is consistent with that estimated from the CN and HCN absorption observed with ALMA at a comparable angular resolution of 1$\arcsec$ \citep{ALMA15}.
A consistent rotational profile was measured in the $^{13}$CO emission, $(2.36\pm0.12) \times (R_{\rm rot}/R_0)^{-0.56\pm0.06}$ km s$^{-1}$ with $V_{\rm sys} = 7.17\pm0.03$ km s$^{-1}$, although the measured $V_{\rm sys}$ in the $^{13}$CO emission is larger than that in the C$^{18}$O emission. 
If the $V_{\rm sys}$ is fixed at $V_{\rm LSR} = 7.04$ km s$^{-1}$, 
the rotational profile in the $^{13}$CO emission is measured to be $(2.47\pm0.12) \times (R_{\rm rot}/R_0)^{-0.52\pm0.05}$ km s$^{-1}$ (Fig.~\ref{pvmaj}e). 
Here the rotational velocity is not yet corrected for the inclination.
Our results show that the C$^{18}$O and $^{13}$CO emission lines in the high-velocity ranges trace the same gas motion.

The $^{13}$CO emission is optically thicker than the C$^{18}$O emission and is expected to trace upper layers of the disk, where the rotational velocity is lower than that in the midplane at the same radius. 
Nevertheless, the rotational profiles measured with these two lines are consistent within the uncertainties, suggesting that the effect of the scale height on our measurements is negligible with our sensitivity and resolutions. 
In addition, we note that the measured $V_{\rm sys}$ with these two lines are different by 0.1 km s$^{-1}$. 
The $V_{\rm sys}$ measured with the $^{13}$CO line is biased to a higher value because of the three data points at the highest redshifted velocities ($>$5 km s$^{-1}$). 
These high-velocity data points are not detected in the C$^{18}$O emission, which is likely due to the lower optical depth and thus fainter emission of the C$^{18}$O line compared to the $^{13}$CO line. 
The three data points deviates from the overall rotational profile. 
Their origin is not clear, and their velocity structure is not resolved, i.e., the data points at different velocities show almost consistent radii. 
We have tested that if those three data points are removed in the analysis, the $V_{\rm sys}$ measured with the $^{13}$CO line is consistent with that with the C$^{18}$O line within the uncertainty. 
Nevertheless, the overall rotational profile is not affected by these data points.

From the measured rotational profiles, the protostellar mass of \object{HL~Tau} is estimated to be 1.8 $M_\odot$, on the assumption that the inclination angle of the protoplanetary disk is 47$\degr$ \citep{ALMA15}. 
As discussed in \citet{Yen13} and \citet{Aso15}, 
the method has a systematic error of 10\%--20\% due to the limited resolutions and the possible contamination from other motions, 
so the uncertainty of our measured protostellar mass is estimated to be 0.3 $M_\odot$. 
In addition, changing the inclination angle by $\pm$5$\degr$ results in a change in the measured protostellar mass by $^{+0.4}_{-0.2}$ $M_\odot$. 
Our measured protostellar mass is consistent with 1.7 $M_\sun$ estimated from the P--V diagrams in the HCO$^+$ (1--0) and CO (1--0) emission observed with ALMA at a comparable angular resolution of $\sim$1$\arcsec$ by \citet{Pinte16}.

To estimate the disk mass, we fitted a two-dimensional Gaussian function to the central C$^{18}$O emission in the moment 0 map (Fig.~\ref{mom}). 
The total flux is measured to be 2.26$\pm$0.22 Jy km s$^{-1}$ with the de-convolved FWHM size of 1\farcs87$\pm$0\farcs12 $\times$ 1\farcs28$\pm$0\farcs08 (262$\pm$17 au $\times$ 179$\pm$11 au).
The gas kinematic temperature in the protoplanetary disk around \object{HL~Tau} was estimated to be 60 K at the outer radii of 60--100 au from the ALMA observations in the CO (1--0) emission \citep{Yen16}. 
On the assumption of an LTE condition and optically thin C$^{18}$O emission, 
the mass traced by the C$^{18}$O emission is estimated to be 3.5 $\times$ 10$^{-3}$ $M_\sun$ with an excitation temperature of 60 K and a C$^{18}$O abundance of 1.5 $\times$ 10$^{-7}$ relative to H$_2$ \citep{Brittain05, Smith15}.
If the excitation temperature is assumed to be 100 K \citep{Brittain05}, the estimated mass becomes 5 $\times$ 10$^{-3}$ $M_\sun$.
The gas mass in the protoplanetary disk around \object{HL~Tau} estimated with the C$^{18}$O emission is 0.2\%--0.3\% of the protostellar mass and is comparable to the dust mass estimated with the continuum emission, (0.3--3) $\times$ 10$^{-3}$ $M_\sun$ \citep{D'Alessio97, Men'shchikov99, Kwon11, Pinte16, Carrasco16}.
That could imply a low gas-to-dust mass ratio of $<$10.  
Such a low gas-to-dust mass ratio in the protoplanetary disk around \object{HL~Tau} has been suggested by \citet{Pinte16} by modelling the 2.9 mm, 1.3 mm, and 0.87 mm continuum data obtained with ALMA. 
On the other hand, 
observations have found that CO molecules can be depleted by two orders of magnitude in protoplanetary disks because of carbon depletion \citep[e.g.,]
[]{Schwarz16}. 
If that is the case, 
the C$^{18}$O abundance can be lower than the typical value in ISM even when the temperature is higher than the CO sublimation temperature, 
and the disk mass estimated with the C$^{18}$O emission in \object{HL~Tau} would be higher, resulting in a higher gas-to-dust mass ratio.  
Besides, 
the radiative transfer model suggests that the 1.3 mm continuum emission in the protoplanetary disk around \object{HL~Tau} is optically thick with $\tau$ ranging from 2 to 6 except at radii of the gaps \citep{Pinte16}. 
The gas mass of the disk estimated from our continuum subtracted C$^{18}$O flux can be biased to a lower value because of the continuum optical depth. 
Thus, our estimated disk gas mass should be considered as a lower limit because of the possible CO depletion and the effect of the continuum opacity.

\subsection{Temperature, Column Density, and Mass of arc Structures}\label{tnm}

\begin{figure}
\resizebox{\hsize}{!}{\includegraphics{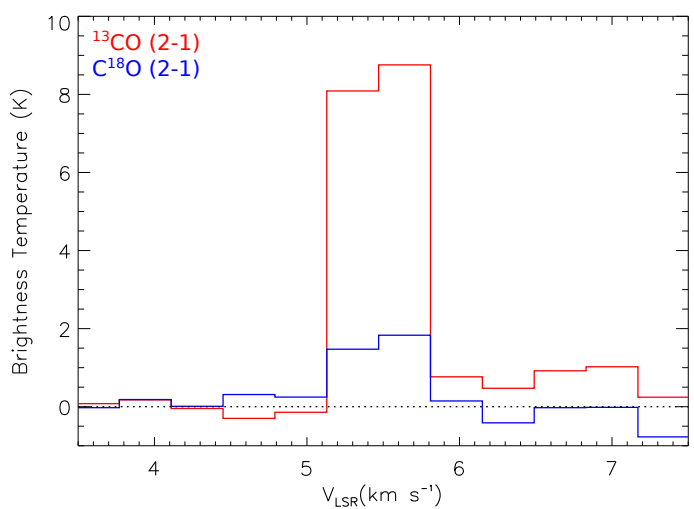}}
\caption{Spectra of the C$^{18}$O (blue histogram) and $^{13}$CO (red histogram) emission averaged over the area within a radius of 0\farcs35 centered at the offset of ($-3\farcs5$, 6\farcs5), that is the northwestern C$^{18}$O component at the medium-low velocity in Fig.~\ref{momc18o}c. The averaged area corresponds to the area of the synthesized beam.}\label{spec}
\end{figure}

In the medium-low velocity range, the C$^{18}$O emission exhibits a blueshifted component located at $\sim$7$\arcsec$ to the northwest of \object{HL~Tau} (Fig.~\ref{momc18o}c). 
This C$^{18}$O component is coincident with the northwestern intensity peak in the blueshifted arc structure in the $^{13}$CO emission in the same velocity range (Fig.~\ref{mom13co}).
On the assumption that this C$^{18}$O component and the coincident $^{13}$CO emission trace the same bulk of gas, 
we estimated the gas kinematic temperature and column density by comparing the C$^{18}$O and $^{13}$CO brightness temperatures.
Figure \ref{spec} presents the spectra of the C$^{18}$O and $^{13}$CO emission averaged over the region within a radius of 0\farcs35 centered at the northwestern C$^{18}$O component. 
The averaged area is the same as the size of the synthesized beam.
The spectra show that both the C$^{18}$O and $^{13}$CO emission lines are detected in two velocity channels, corresponding to a line width of 0.68 km s$^{-1}$.
The mean brightness temperatures of the C$^{18}$O and $^{13}$CO emission over the two velocity channels are measured to be $T_{\rm C^{18}O} = 1.7\pm0.2$ K and $T_{\rm ^{13}CO} = 8.4\pm0.2$ K, respectively. 
The C$^{18}$O (2--1) and $^{13}$CO (2--1) lines are thermalized at typical densities in protostellar sources on a 1000 au scale, where $n({\rm H_2}) > 10^4$ cm$^{-3}$ \citep[e.g.,][]{Shirley00}.
Thus, the excitation temperatures of these two lines are expected to be the same as the gas kinematic temperature. 
The excitation temperature ($T_{\rm ex}$) and their optical depths ($\tau_{\rm C^{18}O}$ and $\tau_{\rm ^{13}CO}$) can be derived from their line ratio as, 
\begin{equation}\label{Tex}
\frac{T_{\rm ^{13}CO}}{T_{\rm C^{18}O}} = (\frac{B_{\nu_{\rm C^{18}O}}(T_{\rm ex}) - B_{\nu_{\rm C^{18}O}}(T_{\rm bg})}{B_{\nu_{\rm ^{13}CO}}(T_{\rm ex}) - B_{\nu_{\rm ^{13}CO}}(T_{\rm bg})})(\frac{1-e^{-\tau_{\rm C^{18}O}}}{1-e^{-\tau_{\rm ^{13}CO}}}), 
\end{equation}
where $B_\nu$ is the Planck function at the frequency $\nu$, $\nu_{\rm C^{18}O}$ and $\nu_{\rm ^{13}CO}$ are the rest frequencies of C$^{18}$O (2--1) and $^{13}$CO (2--1), and $T_{\rm bg}$ is the cosmic background temperature of 2.73 K. 
The ratio of $\tau_{\rm C^{18}O}$ to $\tau_{\rm ^{13}CO}$ can be computed with a given $T_{\rm ex}$ and a $^{13}$CO/C$^{18}$O abundance ratio to solve the above equation. 
In our calculation, the $^{13}$CO/C$^{18}$O abundance ratio was adopted to be 10, which is measured from infrared spectra of CO and its isotopologues in \object{HL~Tau} \citep{Brittain05, Smith15}. 
This $^{13}$CO/C$^{18}$O abundance ratio is higher than the typical ratio of 7 in ISM and 5.5 in the solar system \citep{Wilson94}. 
With Eq.~\ref{Tex}, 
$T_{\rm ex}$ is derived to be 15 K, $\tau_{\rm C^{18}O}$ to be 0.18, and $\tau_{\rm ^{13}CO}$ to be 1.78. 
Thus, the gas kinematic temperature at the northwestern peak in the arc structures is estimated to be 15 K. 
Assuming this component is in the disk plane, its de-projected radius is 7\farcs6 (1060 au). 
The estimated gas kinematic temperature is consistent the expected dust temperature at a distance of 1000 au from a star with a luminosity of 7.6 $L_\odot$ (15 K; \citet{Beckwith90, Hayashi93}), and it is lower than the gas temperature of $\sim$60 K in the outer region ($\sim$60--100AU) of the protoplanetary disk around \object{HL~Tau} measured with the ALMA CO (1--0) observations \citep{Yen16}.
If a lower $^{13}$CO/C$^{18}$O abundance ratio of 5.5 is adopted, 
the derived $T_{\rm ex}$, $\tau_{\rm C^{18}O}$, and $\tau_{\rm ^{13}CO}$ become 38 K, 0.05, 0.3, respectively.

Except for the northwestern C$^{18}$O component, 
there is no other C$^{18}$O counterpart of the arc structures clearly detected. 
Thus, to estimate the mass of the arc structures, 
we interpolated the temperature assuming it is a power-law profile and the temperature is 15 K at a radius of 1000 au (this work) and 60 K at a radius of 100 au \citep{Yen16}. 
The power-law index is derived to be $-0.6$. 
Then, we computed the optical depth of the $^{13}$CO emission pixel by pixel from its mean brightness temperature in the medium-velocity ranges on the assumption of an LTE condition. 
Our calculation shows that the $^{13}$CO emission is mostly optically-thin to close to optically thick with $\tau$ ranging from 0.3 to 1, except for the northwestern end of the blueshifted arc structure, where $\tau$ increases to 2--3. 
From Fig.~\ref{mom13co}b and c, the integrated fluxes of the blue- and redshifted arc structures are measured to be 866 Jy km s$^{-1}$ over an area of 194 square arcsecond and 533 Jy km s$^{-1}$ over an area of 110 square arcsecond, respectively. 
To exclude the contribution from the protoplanetary disk at the medium velocities, 
the flux in the central region, where intense C$^{18}$O emission is present (light blue contours in Fig.~\ref{modelv}), was not integrated. 
The masses of the blue- and redshifted arc structures are estimated to be 2.9 $\times$ 10$^{-3}$ $M_\sun$ and 2.8 $\times$ 10$^{-3}$ $M_\sun$ on the assumption of the $^{13}$CO abundance of 1.5 $\times$ 10$^{-6}$ relative to H$_2$ \citep{Brittain05, Smith15}. 
The correction of the optical depth, $\tau/(1-\exp^{-\tau})$, was performed pixel by pixel. 
The estimated mass of the arc structures is comparable to or lower than the mass of the protoplanetary disk estimated from the C$^{18}$O emission (Sect. \ref{kep}) and also from the continuum emission (0.003--0.3 $M_\sun$; e.g., \citet[][]{Pinte16, Carrasco16}). 
In addition, it is approximately one-tenth of the total mass on a scale of 20$\arcsec$ (2800 au) around \object{HL~Tau} measured with the Nobeyama Millimeter Array at an angular resolution of 5$\arcsec$ (0.03 $M_\odot$; \citet{Hayashi93}) and with the IRAM 30-m telescope at an angular resolution of 22$\arcsec$ (0.06 $M_\odot$; \citet{Cabrit96}). 
The interpolated temperature profile shows that the kinematic temperature is higher than 20 K, the CO sublimation temperature, at a radius less than 600 au. 
Hence, CO molecules are likely not frozen out in the arc structures, and thus our mass estimation is not affected by CO depletion, except for the northwestern end of the blueshifted arc structure where the temperature is likely lower than 20 K. 

\subsection{Kinematics of arc Structures}\label{kin}
\begin{figure*}
\centering
\includegraphics[width=17cm]{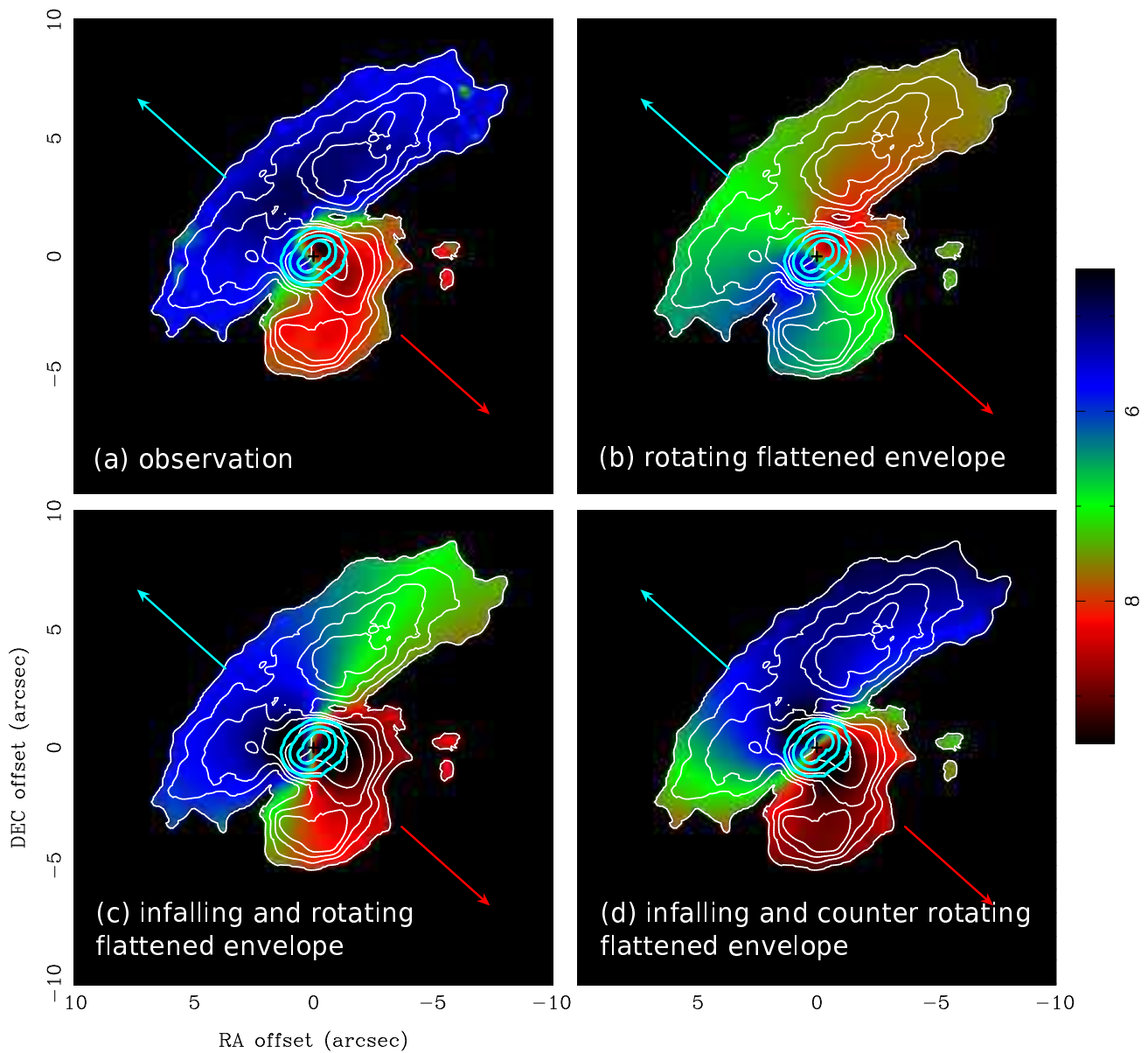}
\caption{White and light blue contours present the moment 0 map of the $^{13}$CO emission at the medium velocities and the C$^{18}$O emission, respectively. Color scales present the moment 1 maps of (a) the observed $^{13}$CO emission computed with the whole velocity range, same as Fig.~\ref{mom} left, (b) Keplerian rotating envelope model, (c) infalling and rotating envelope model, and (d) infalling and counter-rotating envelope model in units of km s$^{-1}$ in the LSR frame. Blue and red arrows show the directions of the blue- and redshifted outflows. Crosses denote the position of \object{HL~Tau}. Contour levels are 5\%, 10\%, 15\%, 20\%, 35\%, 70\%, and 90\% of the peak intensity in the $^{13}$CO map, and are 20\%, 50\%, and 80\% of the peak intensity in the C$^{18}$O map. The peak intensities are 0.78 and 0.41
$^{-1}$ km s$^{-1}$, corresponding to 94$\sigma$ and 52$\sigma$, in the $^{13}$CO and C$^{18}$O maps, respectively.}\label{modelv}
\end{figure*}

Figure \ref{modelv} presents the moment 0 maps of the $^{13}$CO emission integrated over the medium-velocity ranges only and overlaid on the moment 1 map of the $^{13}$CO emission computed with the whole velocity range to emphasize the velocity pattern of the arc structures. 
The light blue contours show the moment 0 map of the C$^{18}$O emission. 
The velocity features inside these C$^{18}$O contours are dominated by the disk component, and thus, they are not discussed here. 
The arc structures exhibit a dominant velocity gradient along the disk's minor axis from the northeast (blueshifted) to the southwest (redshifted). 
In addition, there is a smaller velocity gradient along the disk's major axis, where the northwestern part is more blueshifted than the southeastern part. 
That is more clearly shown in the P--V digram along the major axis (Fig.~\ref{pvmaj}c). 
The southeastern part is centered at the relative velocity of $-1$ km s$^{-1}$, and the northwestern part is distributed at the relative velocity from $-2.5$ km s$^{-1}$ to $-1$ km s$^{-1}$ with respect to $V_{\rm sys} = -7$ km s$^{-1}$.
To study these observed velocity structures of the arc structures, 
we computed model moment 1 maps of different kinds of gas motions and compared them with the observations.

\subsubsection{Keplerian Rotation Model} 
Figure \ref{modelv}b presents the model moment 1 map assuming that the arc structures are Keplerian rotating around \object{HL~Tau} having a mass of 1.8 $M_\odot$. 
We assume that the arc structures are in the disk plane with an inclination angle of 47$\degr$ and a PA of 138$\degr$, the same as the central protoplanetary disk. 
The model moment 1 map shows a clear velocity gradient along the disk's major axis, 
where the northwestern part is redshifted and the southeastern part is blueshifted,  
and there is no velocity gradient along the disk's minor axis.
In our observations, 
the northwestern part of the arc structures is blueshifted, 
and the velocity pattern is not symmetric with respect to the disk's minor axis. 
In addition, a velocity gradient along the disk's minor axis is clearly observed.
These features are inconsistent with the expectation from a dominant rotational motion, 
and thus the motion of the arc structures is not pure rotation around \object{HL~Tau}. 

\subsubsection{Infall and Rotation Model}  
The arc structures exhibit a clear velocity gradient along the disk's minor axis (northeast--southwest direction), 
which was interpreted as infalling and rotational motions in a disk-like structure by \citet{Hayashi93}. 
To compare our observations with this interpretation, 
we computed models of an infalling and rotating geometrically-thin envelope with our measured protostellar mass and specific angular momentum (see Appendix \ref{kmodel1}). 
The moment 1 map of this model shows a clear velocity gradient along the disk's minor axis similar to the observations (Fig.~\ref{modelv}c), 
suggesting that the observed velocity gradient along the disk's minor axis can be caused by infalling and rotational motions. 
However, in the model moment 1 map, the velocity in the northwestern region is close to the systemic velocity because the velocity of the infalling motion in the disk plane projected onto the line of sight approaches zero near the disk's major axis.
In addition, the rotational motion induces redshifted emission in the northwest, as demonstrated in Fig.~\ref{modelv}b.
As a result, there is no blueshifted velocity in the northwest in the model moment 1 map. 
That is inconsistent with the observations that show clear blueshifted emission in the northwest. 
Therefore, the infalling and rotational motions in the disk plane cannot fully explain the gas motions of the arc structures. 

\subsubsection{Infall and Counter-Rotation Model} 
Alternatively, if the arc structures are infalling and rotating in a counter direction with respect to the disk rotation, 
this counter rotation can induce blueshifted emission toward the northwest and redshifted emission toward the southeast. 
To compare this scenario with the observations, we computed models of an infalling and counter-rotating geometrically-thin envelope (see Appendix \ref{kmodel2}).
The moment 1 map of this model shows that the counter rotation indeed induces blueshifted emission in the northwest, similar to the observations (Fig.~\ref{modelv}d). 
However, the velocity of the eastern part of the arc structures in the model moment 1 map is close to the systemic velocity and even become redshifted. 
That is because the counter rotation induces redshifted emission in the southeast and compensates blueshifted emission induced by the infalling motion. 
This feature is different from the observations where the eastern part of the arc structures is clearly blueshifted. 
Therefore, the gas motions of the arc structures also cannot be fully explained with our model of the infalling and counter rotational motions following a simple rotational profile. 

\begin{figure*}
\centering
\includegraphics[width=17cm]{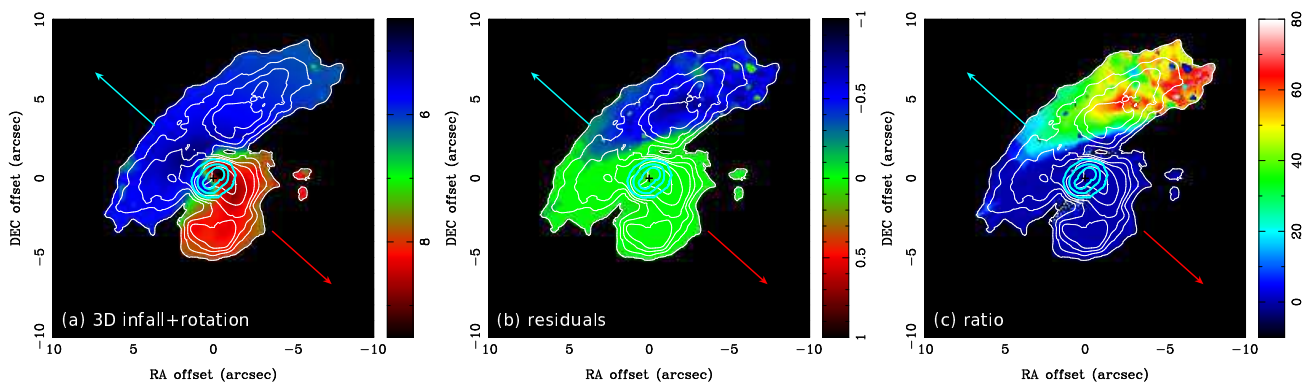}
\caption{(a) Model moment 1 map (color) that best matches the observations, assuming the gas kinematics traced by the $^{13}$CO emission is the three-dimensional infalling and rotational motions as described by Eq.~\ref{vr}--\ref{vlos}. (b) Residual moment 1 map (color) after subtracting the model moment 1 map, panel (a), from the observed moment 1 map, Fig.~\ref{modelv}a. (c) Ratio of the residuals to the expected relative velocity from the model. The color scale shows the percentages. White and light blue contours present the observed moment 0 maps of the $^{13}$CO emission at the medium velocities and the C$^{18}$O emission, the same as those in Fig.~\ref{modelv}. Blue and red arrows show the directions of the blue- and redshifted outflows. Crosses denote the position of \object{HL~Tau}.}\label{3dvel}
\end{figure*}

\subsubsection{Three-Dimensional Infall and Rotation Model} 
We then considered three-dimensional infalling motion because infalling gas above the disk plane can have a non-zero line-of-sight velocity ($V_{\rm los}$) along the disk's major axis after projection, different from infalling motion with the geometrically-thin approximation.
In this model, we adopted the velocity profiles of three-dimensional infalling and rotational motions from \citet{Ulrich76} and \citet{Mendoza04}. 
To take all the possible orientations of the arc structures in the three-dimensional space into account,  
we assume that the arc structures are composed of an assemble of infalling and rotating particles, 
and each particle can be located anywhere along the line of sight.
In this case, all the possible angles between the velocity vectors and a given line of sight were computed. 
For each pixel, we searched for model $V_{\rm LSR}$ that best matches the observations. 
The details of the model calculations are described in Appendix \ref{kmodel3}.

The model moment 1 map indeed shows similar velocity features to the observations, 
suggesting the kinematics of the arc structures can be explained with the three-dimensional infalling and rotational motions, 
where the blue- and redshifted arc structures are located behind and in front of the disk plane, respectively. 
Figure \ref{3dvel}b presents the residuals after subtracting the model from the observed moment 1 map.
The residual map shows that the observed velocity of the redshifted arc structure and the inner and eastern parts of the blueshifted arc structure can be reproduced with the model. 
However, the observed velocity in the northwestern part of the blueshifted arc structure is clearly more blueshifted than the model. 
The blueshifted velocity excess is as large as 0.8 km s$^{-1}$, significantly larger than the velocity resolution of 0.34 km s$^{-1}$. 
In other words, the northwestern part of the blueshifted arc structure has a relative velocity larger than the expected free-fall velocity. 
Note that this velocity excess is the difference between the observed and model $V_{\rm LSR}$ and is not yet corrected for the projection.
Figure \ref{3dvel}c shows the ratio of the residuals to the expected relative velocity (i.e., $V_{\rm los}$ in Eq.~\ref{vlos}) from the model.
Considering the projection, the observed velocity excess is estimated to be $\sim$60\%--70\% of the expected free-fall velocity, which is significantly larger than the uncertainty in the protostellar mass.
As all the possible angles between the velocity vector and the line of sight were considered in our model calculation, 
the deviation between the observed velocity and the expected free-fall velocity is not due to the orientation of the arc structures in the three-dimensional space.
This result suggests that the gravity of \object{HL~Tau} is insufficient to accelerate the infalling gas to have the observed velocity. 
Therefore, the kinematics of the whole arc structures cannot be simply interpreted as infalling flows toward the disk with the conventional model. 

\subsubsection{Effects of Missing Flux on Observed Velocity}
We note that our ALMA observations suffer from missing flux, which may bias our observed mean velocity.  
\object{HL~Tau} has been observed with the IRAM 30m telescope in the $^{13}$CO (2--1) and (1--0) emission \citep{Monin96, Cabrit96}. 
The systemic velocity of the large-scale ambient gas on a scale of 2$\arcmin$ (17\,000 au) around \object{HL~Tau} is estimated to be $V_{\rm LSR}$ = 6.3--6.6 km s$^{-1}$, 0.4--0.7 km s$^{-1}$ more blueshifted compared to the systemic velocity of \object{HL~Tau} estimated from the disk rotation. 
We convolved our $^{13}$CO image with the angular resolution of the IRAM 30m observations in the $^{13}$CO (2--1) emission, 11\farcs4, and compared the observed brightness temperature at the position of \object{HL~Tau}.  
In the blueshifted velocity range, the missing flux is $\sim$40\% at $V_{\rm LSR} = 4.6$ km s$^{-1}$ and increases to 80\% at $V_{\rm LSR} = 5.6$ km s$^{-1}$. 
The missing flux is the highest and more than 95\% at the systemic velocity of the ambient gas, $V_{\rm LSR}$ = 6.3--6.5 km s$^{-1}$. 
At the systemic velocity of the protoplanetary disk around \object{HL~Tau}, $V_{\rm LSR}$ = 7 km s$^{-1}$, the missing flux is also more than 95\%. 
In contrast, in the redshifted velocity range, there is almost no missing flux at $V_{\rm LSR} > 8.4$ km s$^{-1}$. 
If the large-scale ambient gas is optically thick at the low veloities of $V_{\rm LSR} \sim 5\mbox{--}7$ km s$^{-1}$, 
our ALMA observations cannot detect the $^{13}$CO emission (if any) originating from the arc structures at these low velocities and would resolve out the optically-thick surface of the large-scale ambient gas. 
Missing the emission at these low velocities can bias the observed mean velocity in the blueshifted arc structure to be more blueshifted.
In Fig.~\ref{13cospec}, we denote the model velocity of the three-dimensional infalling and rotational motions at these positions. 
If the arc structures indeed follow the free-falling and rotational motions and show the higher velocity because of the missing flux, 
the actual spectra without the missing flux should center at these expected velocity from the model.  
That would imply the actual line width of the blueshifted arc structure to be 2 km s$^{-1}$. 
On the other hand, there is less missing flux in the redshifted velocity range. 
The line width of the redshifted arc structure is narrower than 1.4 km s$^{-1}$ at the offset of F.(0\farcs7, $-3\farcs5$), where it is less blended with the low-velocity outflow. 
Thus, the line width of the blueshifted arc structure is not expected to be as wide as 2 km s$^{-1}$, assuming that the blue- and redshifted arc structures have the same origins and similar physical conditions. 
The high observed mean velocity in the northwestern part of the blueshifted arc structure is unlikely purely caused by the missing flux. 
Nevertheless, adding short-spacing data to fully recover the structures at the low velocities with a high angular resolution is essential to examine this possibility.

\section{Discussion}
Our ALMA observations reveal the arc structures in the $^{13}$CO emission connected to the protoplanetary disk around \object{HL~Tau} (Fig.~\ref{mom13co}b \& c). 
The overall velocity features of the arc structures can be explained with the three-dimensional infalling and rotational motions (Fig.~\ref{3dvel}a). 
However, the observed relative velocity in the northwestern part of the blueshifted arc structure is larger than the expectation from the free-fall motion, 
and the deviation is as large as 0.8 km s$^{-1}$, corresponding to $\sim$60\%--70\% of the expected free-fall velocity (Fig.~\ref{3dvel}).  
Thus, its kinematics cannot be fully explained with the conventional model of infalling and rotational motions \citep[e.g.,][]{Ulrich76}. 
This suggests that the gravity of \object{HL~Tau} is insufficient to accelerate the infalling gas to the observed relative velocity. 
If the observed arc structures are indeed infalling toward the protoplanetary disk around \object{HL~Tau}, 
that requires an additional force besides the gravity of \object{HL~Tau} to drive the infalling motion. 

The $^{13}$CO (1--0) map with a size of 6$\arcmin$ toward \object{HL~Tau} obtained by combining BIMA and NRAO 12m data shows a shell with a size of 2$\arcmin$ $\times$ 1\farcm5 ($\sim$11\,000 au in radius) centered at \object{XZ~Tau} that is located 25$\arcsec$ to the east of \object{HL~Tau} \citep{Welch00}. 
This large-scale shell is interpreted as an expanding bubble with an expanding velocity of 1.2 km s$^{-1}$ based on its velocity structures observed in the $^{13}$CO (1--0) emission, and the expansion is suggested to be driven by \object{XZ~Tau} \citep{Welch00}.
\object{XZ~Tau} is a binary system and classified as a classical T Tauri star. 
Observations with the Hubble Space Telescope reveal several concentric elongated bubbles on a 10$\arcsec$ scale around \object{XZ~Tau}, suggesting \object{XZ~Tau} has episodically ejected wide-angle outflows \citep{Krist99, Krist08}.
The molecular outflows in \object{XZ~Tau} have been observed in the CO (1--0) emission on a 10$\arcsec$ scale with ALMA, 
and the observed velocity features also hint at wide-angle expansion \citep{Zapata15}. 
\object{HL~Tau} is located at the wall of the large-scale expanding shell. 
Thus, the protostellar envelope around \object{HL~Tau} can be compressed by the expansion. 
The single-dish observations indeed show that the systemic velocity of the large-scale ambient gas is different from that of \object{HL~Tau}  \citep{Monin96, Cabrit96}. 
This suggests that the present protostellar envelope around \object{HL~Tau} could have relative motion with respect to \object{HL~Tau}, possibly due to the compression by the expanding shell. 
The dynamical time scale of the expansion can be estimated as the radius of the expanding shell (11\,000 au) over the expanding velocity (1.2 km s$^{-1}$) to be 4 $\times$ 10$^4$ yr. 
That is shorter than the typical life time of the Class 0 and I stage (10$^5$ yr; \citet{Enoch09}).
Thus, initially \object{HL~Tau} could form out of a protostellar envelope having a $V_{\rm sys}$ of 7 km s$^{-1}$, as we measured from the disk rotation (Sect.~\ref{kep}).
Later, the protostellar envelope could be compressed by the expanding shell driven by \object{XZ~Tau} and have motions relative to \object{HL~Tau}, resulting in a different $V_{\rm sys}$, and the outer envelope material could collapse toward \object{HL~Tau} with a non-zero initial velocity.
In this case, the infalling velocity can be higher than the expected free-fall velocity derived from the protostellar mass of \object{HL~Tau}. 
The difference in the observed relative velocity and the expected free-fall velocity in the northwestern part of the arc structures is $\lesssim$0.8 km s$^{-1}$. 
That is comparable to the difference in the systemic velocities between \object{HL~Tau} and the large-scale ambient gas, 0.4--0.7 km s$^{-1}$, and is smaller than the expanding velocity of the shell, 1.2 km s$^{-1}$.
Therefore, the scenario of the infalling envelope externally compressed by the expanding shell is able to explain the observed high velocity of the blueshifted arc structure, 
and the arc structures are infalling toward the protoplanetary disk around \object{HL~Tau}. 

The morphologies of the arc structures observed in \object{HL~Tau} are much more bent than the expected trajectories of material free falling with a specific angular momentum of 1.9 $\times$ 10$^{-3}$ km s$^{-1}$ pc, corresponding to a centrifugal radius of 100 au, toward \object{HL~Tau} (see Fig.~2 in \citet{Tobin12}). 
Thus, the arc morphologies are unlikely caused by the trajectory of the infalling material, unless the infalling material initially posses a much higher specific angular momentum than the central protoplanetary disk around \object{HL~Tau} and loses its angular momentum via mechanisms such as magnetic braking within a 1000 au scale,  making its velocity vector closer to the radial direction afterward. 
This is different from the case in the other Class I protostar L1489, where the morphologies and velocity structures of the infalling flows can be described with the conventional model of free-fall motion with a constant angular momentum \citep{Yen14}. 
On the other hand, 
asymmetric arc structures are often present in numerical simulations of infalling envelopes. 
In magnetohydrodynamic simulations including the effect of magnetic flux being decoupled from material accreted onto stars, 
the magnetic flux can accumulate and form regions with high magnetic pressure non-isotropically. 
These high-pressure regions can resist infalling material. 
Then infalling material flows along the boundary of the high-pressure region and can appear like arc structures (e.g., Fig.~6 in \citet{Zhao11} and Fig.~6 in \citet{ Krasnopolsky12}).
In addition, in the MHD simulations, where the magnetic field and the rotational axis are misaligned, 
the flattened structures that are formed by mass accumulation perpendicular to the magnetic field direction can be twisted by the rotation. 
The twisted flattened structures can be seen as arcs or spirals after projection with certain inclination angles (e.g., Fig.~1 in \citet{Joos12}).
Nevertheless, the magnetic field tends to slow down infall and rotation \citep[e.g.,][]{Li11}, 
The observed arc structures having the velocity higher than the expected free-fall velocity cannot be explained by simply including the magnetic field effects.
Asymmetric arc structures also form in the simulations incorporating turbulence, 
where the density distribution of an infalling envelope is perturbed by turbulence and can exhibit arc structures (e.g., Fig.~2 in \citet{Li14}). 
It is also possible that the morphologies of the arc structures are simply related to the initial density distribution and velocity structures of the infalling material, which are shaped by the expanding bubble driven by \object{XZ~Tau}. 

If the observed arc structures are indeed infalling toward the protoplanetary disk around \object{HL~Tau}, 
the infalling time scale of the arc structures being fully accreted onto the protoplanetary disk can be estimated as $r/V_r$.
With $V_r$ and $r$ derived from the best-match model velocity field (Fig.~\ref{3dvel}a) and ignoring the additional acceleration caused by the external compression for simplicity, 
the mass-weighted mean infalling time scale is estimated to be $\sim$2600 yr. 
The mass of the arc structures is estimated to be 5.7 $\times$ 10$^{-3}$ $M_\sun$ (Sect. \ref{tnm}). 
Then, the mass infalling rate onto the protoplanetary disk around \object{HL~Tau} is estimated to be $5.7 \times 10^{-3}/2600 = 2.2 \times 10^{-6}$ $M_\sun$ yr$^{-1}$.
Note that the mass of the arc structures is only approximately 10\% of the total gas mass observed on a scale of a few thousand au around \object{HL~Tau}. 
The mass accretion rate onto \object{HL~Tau} can be estimated as $L_{\rm bol}R_*/GM_*$, where $R_*$ is the stellar radius of \object{HL~Tau}.
With $L_{\rm bol} = 3.5\mbox{--}15$ $L_\sun$ \citep{Robitaille07}, $M_* = 1.8$ $M_\sun$, and $R_* = 5$ $R_\sun$ \citep{Palla91}, 
the mass accretion rate is estimated to be (0.3--1.4) $\times$ 10$^{-6}$ $M_\sun$ yr$^{-1}$, 
assuming the bolometric luminosity of \object{HL~Tau} fully originates from the gravitational energy released by the accretion. 
These mass infalling and accretion rates are significantly higher than those in solar-mass T Tauri stars having mass accretion rates of 10$^{-7}$--10$^{-9}$ $M_\sun$ yr$^{-1}$ \citep{Beltran16} and are comparable to those of Class 0 and I protostars \citep[e.g.,][]{Yen17}.
The estimated mass infalling rate onto the disk is comparable to or possibly larger than the estimated mass accretion rate onto \object{HL~Tau}.  
Hence, the disk mass could gradually increases,  
and the disk around \object{HL~Tau} may become gravitationally unstable because of the continuous mass infall from the envelope \citep{Hennebelle17, Tomida17}.

Alternatively, the arc structures observed in \object{HL~Tau} can be outflowing rather than infalling. 
The arc structures show a velocity gradient along the disk's minor axis, where the northeastern part is blueshifted and the southwestern part is redshifted. 
The direction of this velocity gradient is the same as that of the outflow in the CO lines observed with SMA and ALMA \citep{Lumbreras14, ALMA15, Klaassen16}. 
If the blue- and redshifted arc structures are located in front of and behind the disk plane, respectively, 
the observed velocity gradient can be explained with outflowing motion. 
The elongations of the arc structures are perpendicular to the outflow direction.  
The opening angle of the CO outflow is measured to be 60$\degr$--90$\degr$ \citep{Klaassen16},  
and the infrared image also shows a V-shaped morphology with an opening angle of 60$\degr$ on a scale of 3$\arcsec$ \citep{Takami07}. 
The opening angle of the outflow is not as wide as the extension of the arc structures. 
In addition, the inner part of the arc structures has a higher velocity than the outer part. 
This velocity structure is different from the one typically observed in molecular outflows, which shows higher velocities in outer parts \citep{Shu91, Shu00, Lee00}. 
Thus, if the arc structures are indeed moving away from \object{HL~Tau}, 
their outflowing motions are unlikely due to the jets or wind launched by the central star-disk system.
Finally, 
hydrodynamical simulations show that if a circumstellar disk is sufficiently massive and become gravitationally unstable, 
the disk can form several arc structures and/or fragment into gas clumps (e.g., Fig.2 \& 3 in \citet{Vorobyov16}, Fig.~2 in \citet{Tomida17}, and Fig.~3 in \citet{Hennebelle17}).
Arc structures similar to those in the simulations have been observed on scales of 500--1000 au around young stellar objects which are undergoing accretion outbursts with coronagraphic polarimetric imaging at near infrared \citep{Liu16}.
In the simulations, some of these gas clumps can be ejected away via their multi-body interaction, 
and the ejected gas clumps and the central disk are linked with arc structures \citep{Vorobyov16}. 
The ejected gas clumps can have velocities of a few times of their escaping velocities, 
and they have masses of tens of Jupiter mass, 
resulting in a high mass loss rate in a short period (0.1--0.2 $M_\sun$ in 0.1 Myrs; \citet{Vorobyov16}). 
If the observed arc structures are gas clumps ejected by the protoplanetary disk around \object{HL~Tau} due to the gravitational instability in the disk, 
it can explain the observed relative velocity that is higher than the expected free-fall velocity. 
In addition, the mass of the observed arc structures is comparable to those of the ejected gas clumps in the simulations.  
 
Indeed, the analysis of the millimeter and submillimeter continuum emission in the protoplanetary disk around \object{HL~Tau} observed with ALMA at angular resolutions of 0\farcs03--0\farcs07 (5--10 au) suggests that the disk can be gravitationally unstable at a radius larger than $\sim$50 au, 
and the gravitational instability is proposed as a possible mechanism to form hypothetical bodies that open gaps in the disk \citep{Akiyama16}.
Nevertheless, the analysis of the gravitational instability in the protoplanetary disk around \object{HL~Tau} depends on the assumed gas-to-dust mass ratio, which was adopted to be 100. 
With a gas-to-dust mass ratio of 100, the disk mass was estimated to be $\sim$0.13 $M_\sun$ (Kwon et al.~2011), which is 7\% of the protostellar mass. 
That implies that the C$^{18}$O abundance is 40 times lower than the typical ISM value based on the ratio between our estimated disk mass from the C$^{18}$O emission and the disk mass of 0.13 $M_\sun$ estimated from the continuum emission. 
The CO abundance lower than the ISM value by two orders of magnitude has been observed in the protoplanetary disk around \object{TW~Hya} \citep{Schwarz16}, 
and possibly also in several other disks \citep[e.g.,][]{Ansdell16}. 
Therefore, it is possible that the observed arc structures are formed by the gravitational instability in the protoplanetary disk and are moving away from \object{HL~Tau}. 
In this case, the formation of the arc structures and the gaps in the protoplanetary disk around \object{HL~Tau} could be linked. 

\section{Summary}
We perform imaging and analyses on our ALMA cycle-3 data of the $^{13}$CO (2--1) and C$^{18}$O (2--1) emission in the candidate of planet formation, \object{HL~Tau}. The aim is to probe the gas kinematics in the protostellar envelope surrounding the protoplanetary disk and to investigate the interaction between the envelope and the disk. Our main results are summarized below.
\begin{enumerate}
\item{The C$^{18}$O emission primarily traces a compact component with a size of 2$\arcsec$ $\times$ 1\farcs5 (280 au $\times$ 210 au) around \object{HL~Tau}. The $^{13}$CO emission also shows a similar central component. The central components are oriented along the major axis of the protoplanetary disk and exhibit dominant velocity gradients along the disk's major axis.}
\item{Two arc structures on a scale of 1000--2000 au connected to the central component are observed in the $^{13}$CO emission. The blueshifted arc structure stretches from the northeast to the northwest, and the redshifted one from southwest to the southeast. The arc structures show overall velocity gradients along both the disk's major and minor axes. The direction of the velocity gradient along the minor axis is consistent with the outflow and the infalling motion, and that along the major axis is opposite to the rotational direction of the protoplanetary disk. The mass of the arc structures is estimated to be 5.7 $\times$ 10$^{-3}$ $M_\sun$.} 
\item{The rotational profile of the central component is measured to be $(2.44\pm0.07) \times (R_{\rm rot}/R_0)^{-0.52\pm0.04}$ km s$^{-1}$ in the C$^{18}$O emission and $(2.47\pm0.12) \times (R_{\rm rot}/R_0)^{-0.52\pm0.05}$ km s$^{-1}$ in the $^{13}$CO emission, and the systemic velocity is measured to be $V_{\rm LSR} = 7.04\pm0.04$ km s$^{-1}$. This rotational profile is consistent with Keplerian rotation ($f = -0.5$) within the uncertainty, suggesting that this central component traces the protoplanetary disk around \object{HL~Tau}. The protostellar mass of \object{HL~Tau} is estimated to be 1.8$\pm$0.3 $M_\sun$ on the assumption that the inclination angle of the protoplanetary disk is 47$\degr$. The gas mass of the disk is estimated to be 3.5 $\times$ 10$^{-3}$ $M_\sun$ with the ISM C$^{18}$O abundance. That is comparable to the dust mass of the disk, (0.3--3) $\times$ 10$^{-3}$ $M_\sun$, estimated in previous studies of the continuum emission, suggesting a low gas-to-dust mass ratio of $<$10 or a low C$^{18}$O abundance of $<$10$^{-8}$ in the protoplanetary disk around \object{HL~Tau}.}
\item{Comparison between the moment 1 maps of the arc structures observed in the $^{13}$CO emission with the models of different gas motions suggests that the kinematics of the arc structures cannot be fully explained with the infalling and (counter-)rotational motions in the flattened envelope around \object{HL~Tau}. The overall velocity structures can be better explained with the three-dimensional infalling and rotational motions, where the blue- and redshifted arc structures are located behind and in front of the disk plane, respectively. However, the observed mean velocity in the northwest part of the blueshifted arc structure is more blueshifted than the expectation from the model, suggesting that the observed velocity is higher than the expected free-fall velocity. The velocity excess is as large as 0.8 km s$^{-1}$ along the line of sight, corresponding to $\sim$60\%--70\% of the expected free-fall velocity.}
\item{If the arc structures are indeed infalling flows toward the protoplanetary disk around \object{HL~Tau}, its infalling velocity higher than the expected free-fall velocity can be explained with the external compression. \object{HL~Tau} is located at the edge of an expanding shell driven by a nearby T Tauri star, \object{XZ~Tau}. The systemic velocity of the ambient gas on a scale of 0.1 pc is $V_{\rm LSR} = 6.3\mbox{--}6.6$ km s$^{-1}$, more blueshifted than that of the central disk. Thus, the protostellar envelope could have relative motion with respect to \object{HL~Tau} because of the external compression by the expanding shell, resulting in the higher infalling velocity. In this scenario, the mass infalling rate onto the protoplanetary disk is estimated to be 2.5 $\times$ 10$^{-6}$ $M_\sun$ yr$^{-1}$, comparable to or larger than the mass accretion rate onto \object{HL~Tau}, (0.3--1.4) $\times$ 10$^{-6}$ $M_\sun$ yr$^{-1}$. Thus, the protoplanetary disk around \object{HL~Tau} could gradually increase its mass.}
\item{Alternatively, the motions of the arc structures can be outflowing instead of infalling, if the blue- and redshifted arc structures are located in front of and behind the disk plane. The previous ALMA observations in the (sub-)millimeter continuum emission toward the protoplanetary disk around \object{HL~Tau} suggest that the outer part ($r > 50$ au) of the disk can be gravitational unstable. The instability can cause a protoplanetary disk to form arc structures and gas clumps with masses of tens of Jupiter mass. Some of the gas clumps can be ejected via multi-body interaction and move away from the disk at velocities of few times of their escaping velocities, and the ejected gas clumps and the disk are connected with arc structures. Thus, the outflowing motion driven by the instability could explain the high velocity in the northwestern part of the blueshifted arc structure.}
\end{enumerate}

\begin{acknowledgements} 
This paper makes use of the following ALMA data: ADS/JAO.ALMA\#2015.1.00551.S. ALMA is a partnership of ESO (representing its member states), NSF (USA) and NINS (Japan), together with NRC (Canada), NSC and ASIAA (Taiwan), and KASI (Republic of Korea), in cooperation with the Republic of Chile. The Joint ALMA Observatory is operated by ESO, auI/NRAO and NAOJ. We thank all the ALMA staff supporting this work. We acknowledge Bo Zhao for fruitful discussion on formation of arc structures in numerical simulations. S.T. acknowledges a grant from the Ministry of Science and Technology (MOST) of Taiwan (MOST 102-2119-M-001-012-MY3), and JSPS KAKENHI Grant Number JP16H07086, in support of this work. N.H. is supported by the grant from MOST (MoST 105-2112-M-001-026). S.M. acknowledges a grant from MOST (MoST 103-2112-M-001-032-MY3).

\end{acknowledgements}

\begin{appendix}

\section{Velocity Channel Maps of the $^{13}$CO emission}\label{13coA}
\begin{figure*}
\centering
\includegraphics[width=17cm]{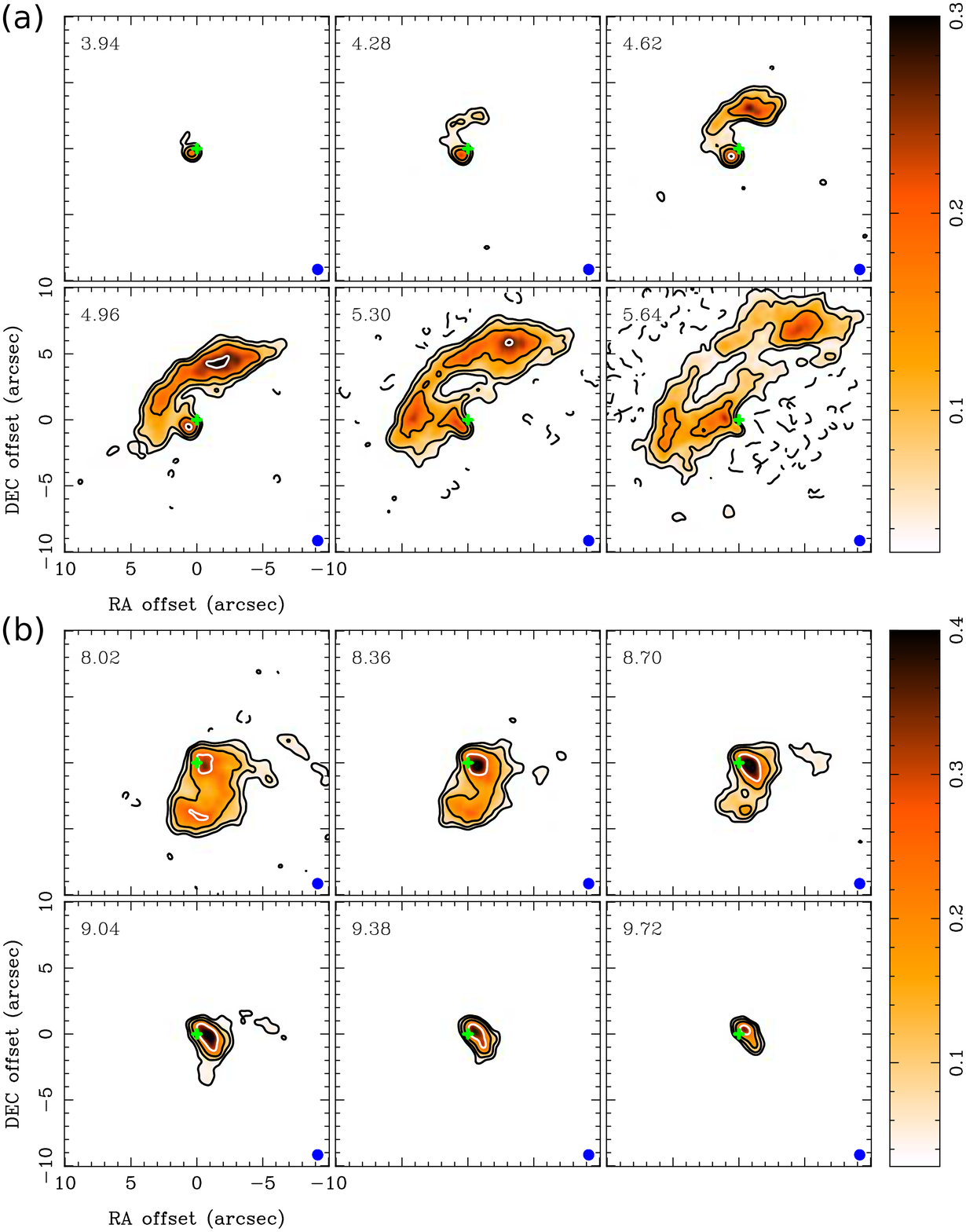}
\caption{Velocity channel maps of the $^{13}$CO emission in the blueshifted (a) and redshifted (b) medium-velocity ranges. Contour levels are 5$\sigma$, 10$\sigma$, 20$\sigma$, and 40$\sigma$, where 1$\sigma$ is 7 mJy Beam$^{-1}$. Color scales are in units of Jy Beam$^{-1}$. Crosses denote the position of \object{HL~Tau}. Filled ellipses present the size of the synthesized beam. The LSR velocity in units of km s$^{-1}$ of each channel is labeled at the upper left corner in each panel.}\label{chan13co}
\end{figure*}

\begin{figure*}
\centering
\includegraphics[width=17cm]{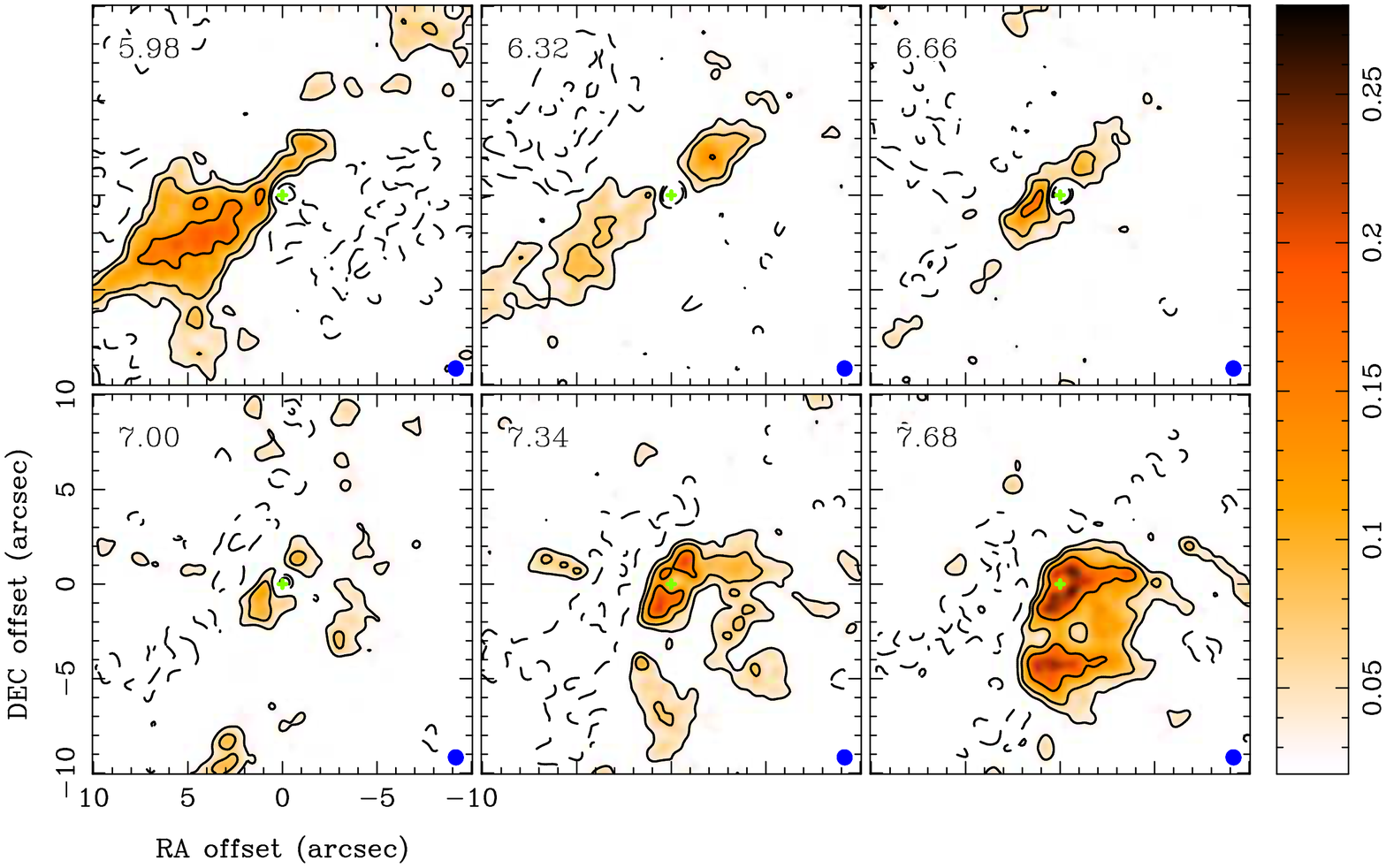}
\caption{Same as Fig.~\ref{chan13co} but for the low-velocity range.}\label{chan13co2}
\end{figure*}

Figure \ref{chan13co} and \ref{chan13co2} presents the whole velocity channel maps of the $^{13}$CO emission at the medium velocities, where the arc structures are observed, and at the low velocities, where the extended structures are observed.

\section{\bf Kinematical Models for the Arc Structures}

\subsection{Infall and Rotation Model}\label{kmodel1}
To compute this model, we adopt the geometrically thin approximation and assume that the arc structures are free falling toward \object{HL~Tau} and rotating with a constant specific angular momentum of 1.9 $\times$ 10$^{-3}$ km s$^{-1}$ pc, the same as that at the outer disk radius. 
The model velocity profiles are adopted to be 
\begin{equation}
V_\phi = V_\phi({\rm 100\ au}) \times (\frac{r}{\rm 100\ au})^{-1},
\end{equation}
\begin{equation}\label{Vin}
V_r = \sqrt{\frac{2GM_*}{r}}, 
\end{equation}
where $V_\phi$ and $V_r$ are the azimuthal and radial velocities, $G$ is the gravitational constant, $M_*$ is the measured protostellar mass of \object{HL~Tau} (1.8 $M_\sun$), and $V_\phi$(100 au) is 3.9 km s$^{-1}$ from the measured rotational profile (Sect. \ref{kep}).

\subsection{Infall and Counter-Rotation Model}\label{kmodel2}
In this model, we also adopt the geometrically thin approximation, 
and we assume that the arc structures are free falling toward \object{HL~Tau} and rotating in the counter direction with respect to the disk rotation.
The observed relative velocity in the northwestern part of the blueshifted arc structure is $\sim$1.5 km s$^{-1}$ at a projected radius of 7$\arcsec$ ($\sim$1000 au; Fig.~\ref{spec}).
If we attribute this relative velocity to the counter rotation, 
it corresponds to a rotational velocity of 2.1 km s$^{-1}$ after correcting the inclination angle with the geometrically thin approximation, 
and it is higher than the expected free-fall velocity at a radius of 1000 au, 1.8 km s$^{-1}$. 
Thus, the arc structures are expected to be rotationally supported if the angular momentum is conserved. 
If the arc structures are indeed infalling and rotating in the counter direction, 
their angular momentum has to be dissipated as they fall toward the center. 
To mimic the angular momentum dissiplation, 
we assume the model velocity profile of the counter rotation to be 
\begin{equation}
V_\phi = V_\phi({\rm 1000\ au}) \times (\frac{r}{\rm 1000\ au}), 
\end{equation}
where $V_\phi({\rm 1000\ au}) = 2.1$ km s$^{-1}$. 
The model velocity profile of the infalling motion is the same as Eq.~\ref{Vin}.

\subsection{Three-Dimensional Infall and Rotation Model}\label{kmodel3}
The model velocity profiles of three-dimensional infalling and rotational motions were adopted from \citet{Ulrich76} and \citet{Mendoza04} as, 
\begin{equation}\label{vr}
V_r = -\sqrt{\frac{GM_*}{r}} \cdot \sqrt{1+\frac{\cos\theta}{\cos\theta_0}},
\end{equation}
\begin{equation}
V_\theta = \sqrt{\frac{GM_*}{r}}\cdot\frac{(\cos\theta_0-\cos\theta)}{\sin\theta}\cdot\sqrt{1+\frac{\cos\theta}{\cos\theta_0}},
\end{equation}
\begin{equation}\label{vphi}
V_\phi = \sqrt{\frac{GM_*}{r}}\cdot\frac{\sin\theta_0}{\sin\theta}\cdot\sqrt{1-\frac{\cos\theta}{\cos\theta_0}},
\end{equation}
where $\theta$ is the angle between the radius of infalling gas and the polar axis, $\theta_0$ is the initial $\theta$ of infalling gas, and $V_\theta$ is the velocity along the $\theta$ direction.
This model describes material free falling toward a point mass with zero total energy and a conserved angular momentum. 
The $V_{\rm los}$ of three-dimensional infalling and rotational motions after projection can be computed as, 
\begin{eqnarray}\label{vlos}
V_{\rm los} = ((V_r\sin\theta-V_\theta\cos\theta)\sin\phi+V_\phi\cos\phi)\sin i \nonumber \\
+ (V_r\cos\theta-V_\theta\sin\theta)\cos i, 
\end{eqnarray}
where $\phi$ is the azimuthal angle. 
The inclination angle and the rotational axis of the three-dimensional model are adopted to be the same as those of the protoplanetary disk around \object{HL~Tau}.

Along a given line of sight, the infalling gas located closer to the disk plane has a larger infalling velocity because of the smaller distance to the star, while its velocity vector is almost on the disk's major axis and is perpendicular to the line of sight. 
The two effects compensate for each other and result in a small $|V_{\rm los}|$.
On the other hand, the infalling gas located further away from the disk plane has a smaller infalling velocity and a velocity vector more parallel to the line of sight. That also results in a small $|V_{\rm los}|$.
Hence, there is a maximum $|V_{\rm los}|$ induced by the infalling and rotational motions along a given light of sight, where the distance of the infalling gas to the disk plane and the angle between its velocity vector and the line of sight are moderate. 
To generate the model moment 1 map of the three-dimensional infalling and rotational motions, 
we assume that the arc structures are composed of an assemble of infalling and rotating particles, 
and each particle can be located anywhere along the line of sight. 
Then, for a given pixel in the model map, we computed all the possible $V_{\rm los}$ of these infalling and rotating particles along the line of sight with Eq.~\ref{vlos}. 
If there is a particle having its $V_{\rm los}+V_{\rm sys}$ ($=V_{\rm LSR}$) that matches the observed velocity at the same position, 
its $V_{\rm los}+V_{\rm sys}$ is assigned to that pixel in the model moment 1 map. 
If none of the computed $V_{\rm los}+V_{\rm sys}$ can match the observed velocity, 
the $V_{\rm los}+V_{\rm sys}$ of the particle having the maximum $|V_{\rm los}|$ along the line of sight is assigned to that pixel. 
In our calculation, $V_{\rm sys}$ is adopted to be 7.04 km s$^{-1}$, the same as that of the central protoplanetary disk.
With this process, we generated a model moment 1 map of the three-dimensional infalling and rotational motions that best match the observations (Fig.~\ref{3dvel}a). 

\end{appendix}

\end{document}